\documentclass[twocolumn]{aastex631}

\usepackage[version=4]{mhchem} 
\usepackage{makecell}
\usepackage{xcolor}
\usepackage{booktabs}
\usepackage{gensymb}
\usepackage{amsmath}
\usepackage{hyperref}

\begin{document}

\title{Investigation of Calcium Dust in the Interstellar Medium}

\author[0009-0002-3333-9249]{Ya-Yu Chu}
\affiliation{Institute of Astronomy and Astrophysics, Academia Sinica, 11F of AS/NTU Astronomy-Mathematics Building, No. 1, Sec. 4, Roosevelt Rd, Taipei 106319, Taiwan}
\affiliation{Department of Physics, National Tsing Hua University, No. 101, Section 2, Kuang-Fu Road, Hsinchu 300013, Taiwan}

\author[0000-0002-8163-8852]{Sascha Zeegers}
\affiliation{Institute of Astronomy and Astrophysics, Academia Sinica, 11F of AS/NTU Astronomy-Mathematics Building, 1, Sec. 4, Roosevelt Rd, Taipei 106319, Taiwan}
\affiliation{SRON Space Research Organisation Netherlands, Niels Bohrweg 4, 2333 CA Leiden, The Netherlands}
\affiliation{Anton Pannekoek Institute for Astronomy, Universiteit van Amsterdam, Science Park 904, 1098 XH Amsterdam, The Netherlands}

\author[0000-0001-8385-9838]{Hsien Shang}
\affiliation{Institute of Astronomy and Astrophysics, Academia Sinica, 11F of AS/NTU Astronomy-Mathematics Building, No. 1, Sec. 4, Roosevelt Rd, Taipei 106319, Taiwan}

\author[0000-0001-8509-0655]{Chi-Liang Chen}
\affiliation{
National Synchrotron Radiation Research Center, No. 101, Hsin-Ann Road, Hsinchu 30092, Taiwan}

\author{Mau-Tsu Tang}
\affiliation{
National Synchrotron Radiation Research Center, No. 101, Hsin-Ann Road, Hsinchu 30092, Taiwan}

\author[0000-0001-9263-8295]{Ming-Jye Wang}
\affiliation{Institute of Astronomy and Astrophysics, Academia Sinica, 11F of AS/NTU Astronomy-Mathematics Building, No. 1, Sec. 4, Roosevelt Rd, Taipei 106319, Taiwan}

\author[0000-0002-1683-3141]{Yun-Liang Soo}
\affiliation{Department of Physics, National Tsing Hua University, No. 101, Section 2, Kuang-Fu Road, Hsinchu 300013, Taiwan}

\author[0000-0001-8470-749X]{Elisa Costantini}
\affiliation{SRON Space Research Organisation Netherlands, Niels Bohrweg 4, 2333 CA Leiden, The Netherlands}
\affiliation{Anton Pannekoek Institute for Astronomy, Universiteit van Amsterdam, Science Park 904, 1098 XH Amsterdam, The Netherlands}

\author[0000-0002-1049-3182]{Ioanna Psaradaki}
\affiliation{European Space Agency (ESA), European Research and Technology Centre (ESTEC), Keplerlaan 1, 2201 AZ Noordwijk, The Netherlands}

\author[0000-0002-5359-9497]{Daniele Rogantini}
\affiliation{Department of Astronomy and Astrophysics, University of Chicago, 5640 S Ellis Ave, Chicago, IL 60637, USA}
\affiliation{MIT Kavli Institute for Astrophysics and Space Research, Massachusetts Institute of Technology, Cambridge, MA 02139, USA}

\correspondingauthor{Ya-Yu Chu;Sascha Zeegers;Hsien Shang}
\email{yychu@asiaa.sinica.edu.tw;\\
S.T.Zeegers@sron.nl;\\
shang@asiaa.sinica.edu.tw}

\begin{abstract} 
Calcium is a highly depleted element in the interstellar medium (ISM), however its composition in the solid phase remains elusive. The detection of Ca-bearing dust has been limited due to its lack of distinct spectral features in the cold phase of the diffuse ISM. X-rays provide a direct method for exploring the absorption and scattering properties of interstellar dust. In this study, we utilize high-resolution X-ray absorption spectroscopy to characterize Ca-bearing dust analogs at the Ca K-edge using X-ray Absorption Fine Structure (XAFS). We present new X-ray absorption spectra of seven calcium-bearing interstellar dust analogs measured at the National Synchrotron Radiation Research Center (NSRRC) in Taiwan. The extinction cross sections, calculated from the laboratory measurements are incorporated in the dust absorption model AMOL of the X-ray spectral fitting tool SPEX.
We select three Low Mass X-ray Binaries (LMXBs) with high column densities, GX 340+00, GX 5-1, and GX 13+1, as background sources to study the intervening ISM along their sightlines. 
The best-fit models are compared with \textit{Chandra} archive data using SPEX.  Additionally, we employ simulations with the \rm{XRISM} and \rm{NewAthena} telescopes to achieve enhanced resolution and more effective coverage at the energy range near the Ca K-edge. These simulations set the boundary conditions for the future X-ray observation of calcium in the interstellar medium.

\end{abstract}

\keywords{Interstellar medium --- Astrochemistry --- Spectroscopy --- X-ray astronomy --- X-ray binary stars}

\section{Introduction} \label{sec:intro}

Interstellar dust plays an important role in the life cycle of stars; calcium is the 13th most abundant element in the solar system and a major element of the upper continental crust in the Earth's crust~\citep{2009LanB...4B..712L, RUDNICK20141}. Calcium resides primarily in the silicate and carbonate phases. It is a fundamental building block for life on Earth and an important lithophile element in terrestrial minerals, such as pyroxene, calcite and dolomite, which allow micro-organisms to live on the fractured surfaces of these rocks~\citep{2014E&PSL.394..135V}. In space, some silicates of interstellar origin contain oldhamite ([Ca,Mg]S)~\citep{Sanghani22}. These grains may form in the atmospheres of evolved stars and may have acted as seed grains for larger silicate dust grains. However, it is still unclear how these grains survive in the harsh conditions of the interstellar medium, where they are bombarded with radiation and particles~\citep{2011Jones}. Regarding calcium silicate hydrate (e.g., $\mathrm{Ca_4Si_4O_{14}H_4}$,  $\mathrm{Ca_6Si_3O_{13}H_2}$, and  $\mathrm{Ca_{12}Si_6O_{26}H_4}$), that is, cementitious particles, it has been proposed that they may exist in an extraterrestrial environment~\citep{Bilalbegovic14}.

Calcium is believed to be produced in significant quantities in the atmospheres of evolved stars, massive stars, AGB stars and released during core-collapse supernovae~\citep{2019A&A...623A.151P,2024Univ...10..148B}; and it can be a fingerprint for supernova remnants~\citep{Draine11, 2025NatAs...9.1356D}. In terms of the origins of calcium-bearing dust, there is one hypothesis that calcium carbonates ($\mathrm{CaCO_3}$) are less favorable than calcium silicates since they need to form in supernova events where the temperature of the environment is tremendously high~\citep{Kempernat02, Ferrarotti05}. In addition, the formation of silicates and oxides are restricted under specific local temperature and total pressure in different environments. Therefore, the chemical structure formed in space is formed under lower pressure so that calcium carbonate will not form into aragonite ($\mathrm{CaCO_3}$) but rather calcite ($\mathrm{CaCO_3}$). In addition, mineral phases such as feldspar group mineral (e.g Anorthite ($\mathrm{CaAl_2Si_2O_8}$)) that form under high pressure are expected to be absent in the interstellar dust~\citep{2022FrASS...9.8217T}.

According to thermodynamic equilibrium models, calcium is among the first elements to condense from the cooling solar nebula at the early stage of disk evolution, transitioning from the gas phase into highly refractory mineral phases~\citep{1998M&PS...33.1123P, 2000GeCoA..64..339E, 1995GeCoA..59.3413Y, 1993GeCoA..57.2377W}. Due to their similar condensation temperatures (1400-1600 K), calcium and aluminum are found to condense together to form refractory solids~\citep{1974RvGSP..12...71G}. These pristine phases are resistant to subsequent thermal and chemical processing during the evolution of the early solar system and are most notably preserved within calcium-aluminum-rich inclusions (CAIs) in carbonaceous chondrites~\citep[][chapter 7]{2021cosm.book.....M}. Consequently, CAIs represent the primary mineralogical records of the transition of Calcium from gas to solid dust. Common CAI phases include diopside ($\mathrm{CaMgSi_2O_6}$) and melilite ($\mathrm{(Ca_2(Mg,Al)(Si,Al)_2O_7)}$), a solid solution of gehlenite ($\mathrm{Ca_2Al_2SiO_7}$) and åkermanite ($\mathrm{Ca_2MgSi_2O_7}$). This mineralogy provides a direct link between interstellar dust and the formation of the early solar system.

The elemental abundance of calcium in the solar system is calculated using CI-type chondrites and is estimated to be $60400$ atoms per $10^6$ Si~\citep{2009LanB...4B..712L}. 
This serves as a benchmark for the expected concentration in the interstellar medium.

Observation of the gas-phase reveals that calcium exhibits high depletion in interstellar space, with a reported mean depletion of $-3.11 \pm 0.55$ dex~\citep{1994ApJ...424..748C}. 
Ca II H and K lines are commonly used as a tracer of this depletion in combination with observations of \citet{1994ApJ...424..748C}. 
Significant fractions of gaseous Ca I and Ca III also exist in the ISM, but are often difficult to measure. To avoid underestimating the gas contribution, the depletion can be compared to other elements, such as Ti II or Na I, from which the unobserved ion stages can be deduced~\citep{Phillips1984,1994ApJ...424..748C, Welty1996}. The depletion measurements of Ca II trace mostly the diffuse warm inter cloud media in both the halo~\citep{Edgar1989} and disk of the Galaxy\citep{1994ApJ...424..748C}. Calcium in denser and colder regions has been explored by~\citet{Cardelli1991}. They find that the strong depletion of calcium in these environments scales with the cloud density.   
Consequently, the majority of calcium is believed to be found in the solid phase in the cold diffuse interstellar medium.

Nevertheless, it is challenging to observe calcium in the interstellar medium (ISM) due to the lack of spectral features in the cold phase. Ongoing efforts across different wavelengths have investigated the presence of calcium features in the ISM; for instance, \citet{Kempernat02} and \citet{Ferrarotti05} analyzed a range of calcium bearing materials using infrared spectra. In the X-rays, several studies of the Fe L, O K, Mg K and Si K-edge provided constraints on the composition of silicates. However, the models used in these studies did not contain calcium silicate species~\citep[see][and references therein]{refId0Si, 2019A&A...630A.143R,Psaradaki2020}.

While calcium is not among the most abundant elements, its highly refractory nature makes it an essential building block and a primary tracer of interstellar dust. Direct X-ray spectroscopy along the interstellar line of sight is thus a powerful tool for this study, utilizing the high penetrative power of X-rays to probe the total column density of these materials. From the X-ray absorption spectrum, we obtain X-ray Absorption Fine Structure (XAFS) features, which arise from the interaction between the photoelectron and the local atomic environment of the dust particle~\citep{article}.

Because XAFS serves as a unique spectral fingerprint for specific mineralogies and crystalline structures, it allows for the identification of the dominant calcium bearing species. Determining whether calcium is incorporated into silicates, carbonates, or sulfides, or exists as metallic calcium, thus helps characterize the chemical composition and mineralogical state of dust in the ISM.

As shown in \citet{Costantini19}, the extinction feature at the calcium K-edge is sensitive to sufficiently high column densities. Although chemical characterization of Ca remains challenging, simulations for \rm{XRISM} and \rm{NewAthena} demonstrate that such features will be detectable. Furthermore, distinctions between calcium bound in carbonate dust versus silicate are suggested to be feasible. The precision of dust size distribution analysis for calcium, as a heavily depleted element, can be determined and may vary depending on the characteristics of the instrument \citep{Costantini19}.

Previous pilot studies of different absorption edges have demonstrated the feasibility of probing dust absorption in the X-ray regime using literature spectra and laboratory spectra measured using synchrotron-based facilities across different atomic absorption edges~\citep[i.e.][]{2005ApJ...622..970L,Costantini19,2020A&A...641A.149R}. 
Building on these studies, it has become a standard approach to integrate laboratory X-ray absorption spectra with astronomical data and measurement in X-rays were also implemented. As they are highly penetrative and allow for the characterization of dust in regions where high optical depth shields the signals typically observed in the infrared. In this article, we present a set of seven dust analogs measured at TPS32A and TLS16A1 beamlines at National Synchroton Radiation Research Center (NSRRC) in Taiwan.

\rm{XRISM} telescope observatory~\citep{Tashiro2025} was launched in September 2023 and provides an unprecedented spectral resolution and a significantly larger effective area through the \texttt{Resolve} micro-calorimeter. \texttt{Resolve} provides a state-of-the-art spectral resolution of 4.5 eV FWHM across the available X-ray energy between 1.7-12 keV, which is a substantial improvement compared to previous CCD-based observatories such as \textit{Chandra}~\citep{2003ExA....16....1W}. Although the \rm{XRISM} gate valve remains closed and restricts the sensitivity below 2 keV, it remains highly capable and optimized for high-precision studies of the Ca K-edge. A recent study using \rm{XRISM} to probe the sulfur K-edge has demonstrated that \rm{XRISM} is a powerful tool for us to investigate the Ca K-edge~\citep{2025PASJ...77S.107C}. In addition, the future X-ray observatory mission, \rm{NewAthena}, is planned to launch in late 2030s. Its calorimeter, X-IFU, has even larger effective area at the Ca K-edge than that from \rm{XRISM}/\texttt{Resolve}~\citep{2025ExA....59...18P,2022arXiv220205399X}.

This study is organized as follows: Section~\ref{sec:laboratory_measurements} details the selection of calcium-bearing mineral samples and the laboratory measurement procedures. Section~\ref{sec:X-ray_source_selection_and_spectral_modeling} describes the selected LMXBs and the spectral fitting and simulation process. Section~\ref{sec:discussion} analyzes the results derived from these simulations. And section~\ref{sec:summary} summarizes the key findings and provides concluding remarks.

\section{Laboratory Measurements of Ca-bearing dust}
\label{sec:laboratory_measurements}
In this section, we describe the laboratory measurements of Ca-bearing dust analogs and the data processing steps. We first obtained the X-ray absorption spectra of selected Calcium-bearing minerals using the tender X-ray absorption spectroscopy TPS32A beamline at the National Synchrotron Research Radiation Center (NSRRC) in Taiwan. The raw measurement data were processed with the \texttt{ATHENA} software \citep{Ravel:ph5155}\footnote{\url{https://bruceravel.github.io/demeter/documents/Athena/index.html}} for spectral normalization around the Ca K-edge energy at 4.038 keV.

From the calibrated data, we derive the optical constants and complex refractive indices. These laboratory-based extinction cross sections were then implemented in the AMOL interstellar dust absorption model in SPEX software for X-ray spectral modeling \citep{kaastra1996_spex, 2024zndo..10822753K}, version 3.08.01. Here we explain this process step by step. 

\subsection{Samples and Dust analogs Collection}
The selection of dust analogs for this study was based on observational evidence from previous literature \citep{2005ASPC..341..605T}. The collection of dust analogs includes Ca-bearing minerals in the silicates and oxides phases. In addition, other phases such as calcium titanate ($\mathrm{CaTiO}_3$), calcium carbonate ($\mathrm{CaCO}_3$) and calcium sulfide ($\mathrm{CaS}$) are also included. Table \ref{tab:samples} provides a comprehensive list of these samples, including their chemical formulas and physical states.

Calcium silicates were selected due to their high predicted fractional abundance, as silicates represent a significant mass fraction of interstellar dust~\citep[e.g.,][]{Henning2010}. Here we include calcium silicate ($\mathrm{CaSiO_{3}}$) and diopside in crystalline and amorphous form, denoted as  $\mathrm{CaMgSi_2O_{6}(cr.)}$ and $\mathrm{CaMgSi_2O_{6}(am.)}$. The amorphous and crystalline diopside samples were synthesized in the Jena laboratory in 1996 and 1998 respectively~\citep{2007ApJ...656..615P}. The amorphous sample was prepared by melting and quenching, a processing technique where the material is heated until molten and then cooled rapidly to form an amorphous solid. The crystalline sample was first mentioned in a study of CAIs by~\citet{2007ApJ...656..615P}, where it was used as a reference material. The calcium silicate is a commercial sample obtained from ThermoFisher. Additionally, calcium titanate was included as it shares the same stoichiometric composition as perovskite, which is a high-temperature condensate and a primary component of the CAIs.

Mineral phases representing different chemical environments were also included. Calcium carbonate ($\mathrm{CaCO_3}$) is generally regarded as a product of aqueous alteration in meteorites~\citep{1992GeCoA..56.2873M,BREARLEY2003247,2025SSRv..221...11L}. However, detections of carbonates in planetary nebulae~\citep{Kempernat02} demonstrate that in the interstellar medium, carbonates can also form through non-aqueous mechanisms such as direct condensation or grain-surface reactions. Moreover, a metallic calcium sample was utilized as a reference standard for the study and to present the possible presence of metallic calcium in the ISM. 

In addition, calcium sulfide ($\mathrm{CaS}$) is found to form in environments with low oxygen fugacity. Its naturally occurring mineral counterpart, oldhamite ($\mathrm{CaS}$), has been identified in enstatite meteorites and formed in sulfur-rich environments~\citep{Sanghani22}. A recent study also found the presence of oldhamite ($\mathrm{CaS}$) on the Lunar regolith~\citep{Li2025}. However, $\mathrm{CaS}$ oxidizes very quickly on Earth when exposed to oxygen in the atmosphere, but this is unlikely to happen in the reduced environment in Space. We noticed a shift in the chemical composition from the measurement at TPS32A, likely due to the long exposure to the atmosphere after the sample has been opened. Therefore, we include the former measurement of $\mathrm{CaS}$ sample at TSL16A performed in fluorescence mode.

\begin{deluxetable*}{clllll}
    \tabletypesize{\footnotesize}
    \tablecaption{Lists of samples measured in this study. \label{tab:samples}}
    \tablecolumns{4}
    \tablewidth{\textwidth}
    \tablehead{ \colhead{ID} & \colhead{Sample Name} & \colhead{Chemical Formula} & molar mass (g/mol) &  density (g/cm$^3$) & \colhead{Remarks}} 
    \startdata 
    1 & Calcium silicate  & \ce{CaSiO3}         & 116.1 &  2.80 & Crystalline \\
2 & Diopside\tablenotemark{a} & \ce{CaMg(SiO3)2}& 216.55 & 3.30 & Crystalline (synth.) \\
3 & Diopside\tablenotemark{a} & \ce{CaMg(SiO3)2}& 216.55 & 3.30 & Amorphous (synth.) \\
4 & Calcium granule   & \ce{Ca}               & 40.08 & 1.53 & Metal piece (crystalline) \\
5 & Calcium titanate  & \ce{CaTiO3}           & 135.94 & 4.10 & Crystalline \\
6 & Calcium carbonate & \ce{CaCO3}            & 100.09 & 2.72 & Crystalline \\
7 & Calcium sulfide   & \ce{CaS}              & 72.14 & 2.59 & Crystalline \\
    \enddata
    \tablenotetext{a}{Samples 2 and 3 were synthesized at the Jena laboratory.\hfill}
    \tablecomments{Sample 1 and 4-7 were obtained through ThermoFisher Scientific (formerly Alfa Aesar).\hfill}
\end{deluxetable*}

\subsection{Laboratory Measurement}
\subsubsection{Principle of the detection method for X-ray absorption spectroscopy}
We utilized Total Electron Yield (TEY) and fluorescence (Fluo) detection to characterize the X-ray absorption properties of the synthesized dust analogs. Unlike transmission mode, which requires stringent control over sample thickness, TEY and Fluo detection are generally less restrictive to the sample thickness.

The TEY method measures the secondary electron current generated at the sample surface. Given the high homogeneity of our synthesized dust analogs, TEY was selected as the primary detection method to avoid self-absorption artifacts. The absorption coefficient, $\mu(E)$, is proportional to the ratio of the measured surface current, $I_e$, to the incident light intensity, $I_0$:
\begin{equation}
\mu(E) \propto \frac{I_e}{I_0}
\end{equation}

For the calcium sulfide (CaS) sample, the measurement was performed using the Fluo detection. To mitigate self-absorption effects inherent to the fluorescence mode, the sample was prepared to be optically thin; a sample thickness of 5 micron was assumed for the calculation of the absorption coefficient and a self-absorption correction was not applied. The absorption coefficient in this mode is defined by the ratio of the fluorescence intensity, $I_f$, to the incident light intensity, $I_0$:
\begin{equation}
\mu(E) \propto \frac{I_f}{I_0}
\end{equation}

\subsubsection{TPS 32A with total electron yield (TEY) detection method}
The beamline TPS 32A is set up using a bending magnet to generate a synchrotron source and provides an energy range from 1.7-11 keV and the energy resolution is estimated to be 0.15 to 1.5 eV with a Si(111) crystal. The instrumental energy resolution ($\Delta E$) is estimated to be approximately 0.48 eV at the Ca K-edge based on beamline optical specifications~\citep{2022JPhCS2380a2041L}, while the absolute energy scale uncertainty is maintained within $\pm 0.2$ eV via concurrent calibration against standard reference materials. The energy calibration was performed prior to spectral acquisition, to ensure the incident energy is aligned at Ca K-edge. We used a Titanium foil at the Ti K-edge (4.966 keV) to establish a reliable relative energy scale. This proxy was selected due to the absence of a metallic Calcium standard foil and the potential oxidation of bulk Calcium pieces from the collection of samples. The absolute absorption energy of Ca K-edge was then aligned and calibrated to the tabulated value of 4.038 keV \citep{thompson2009x}. To improve efficiency, the beamline employs two back-to-back double monochromator configurations that enable fly-scan operation \citep{Liang_2025}. This high-speed mode enables rapid data collection, with a single spectrum typically acquired in approximately 1 minute; however, total measurement time is often extended to enhance the signal-to-noise ratio across the full XANES range. Despite these necessary adjustments, this approach remains significantly more efficient than conventional step-scanning procedures, which require 30–50 minutes for the same energy range.

For the measurement, the procedure is conducted as follows: First, dust samples are taped onto an acrylic disk and measured using the fly scanning method with total electron yield (TEY) detection mode. The samples were quickly prepared to avoid long exposure to air, and the calcium sulfide sample was prepared in a glovebox to avoid moisture and exposure to oxygen. The sample was measured at the 2nd endstation in vacuum around $1 \times 10^{-6}$ torr. This detection method is more time-efficient, and given the fact that TEY detection is surface sensitive, we can rule out the possible self-absorption effect from using fluorescence detection. 

From the TEY detection, we obtain raw absorption spectra from 3.95 to 4.2 keV for Ca, $\mathrm{CaTiO}_3$, $\mathrm{CaCO}_3$ and CaS, and 3.838 to 4.395 keV for $\mathrm{CaSiO}_3$ and $\mathrm{CaMg}(\mathrm{SiO}_3)_2$ in both crystalline and amorphous states.

\subsubsection{TLS 16A1 with fluorescence detection method }
The TLS16A1 provides X-ray absorption spectroscopy measurements in both transmission and fluorescence modes and is equipped with a double crystal monochromator (DCM) for energy selection. For this study, the absorption spectrum of CaS was measured in fluorescence mode. The energy resolving power is $\sim7000$ across the available energy bandpass. 

The CaS sample was prepared in a glove box filled with argon gas. The powder was scooped onto Kapton tape and sealed with Mylar wrap to prevent atmospheric exposure and oxidation. The measurement was conducted in situ, and XAS data was measured using the fluorescence method from 3.8 to 4.6 keV. Initial energy calibration was performed using a Ti foil, and the theoretical energy of Ca K-edge is then set at 4.038 keV~\citep{thompson2009x}. The sample was positioned at an incident angle of $45~\degree$ to facilitate self-absorption correction using the FLUO algorithm within the \texttt{ATHENA} software. This correction accounts for a 10\% uncertainty in the corrected absorption coefficient~\citep{PhysRevB.46.3283}.

Two artifact glitches were observed at 4.02 keV and 4.5 keV resulting from to the arrangement of the mirror angle. These features were removed using the de-glitch function in the \texttt{ATHENA} software. There exists inevitable discrepancy in the shift of the absorption edges between the TLS16A1 and TPS32A beamlines. To maintain the consistency of the spectra across this study, we shift the CaS absorption edge measured at TLS16A1 in fluorescence method to align with the absorption edges at TLS32A toward higher energy by 1.85 eV. The shift is within the instrumental resolution of X-ray telescopes and does not impact or compromise the data application.

\subsubsection{Data Reduction and Processing}
\label{sec:data_processing}
\begin{figure}
    \centering
    \includegraphics[width=1.0\linewidth]{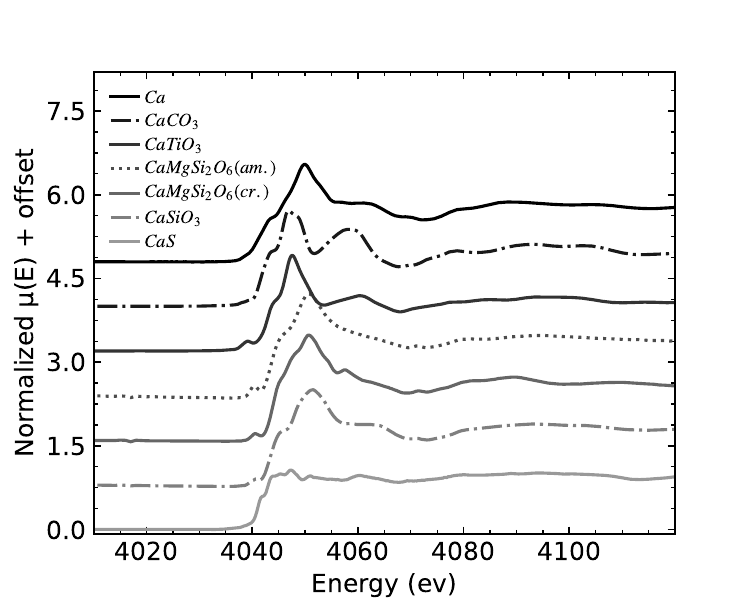}
    \caption{X-ray absorption spectra of the Ca-bearing dust analogues measured at TPS32A and TLS16A1, NSRRC. The y-axis is plotted in normalized absorption coefficient $\mu(E)$.}
    \label{fig:lab_xas_spectra_of_all_dust}
\end{figure}

X-ray Absorption Spectroscopy (XAS) data reduction was performed using the \texttt{ATHENA} software package. The spectra were normalized by fitting second- or third-order polynomials to the pre-edge and post-edge regions to correct for incident beam flux variations and isolate the fine structure by removing the atomic absorption background; no further data calibration was applied. The resulting normalized absorption spectra for the dust analogs are presented in Figure \ref{fig:lab_xas_spectra_of_all_dust}.

\begin{figure}[htb!]
\centering
\begin{minipage}{0.45\textwidth}
  \centering
  \includegraphics[width=1.0\linewidth]{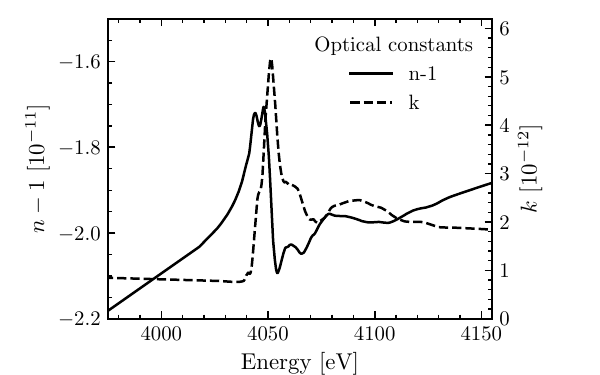}
  (a) 
\end{minipage}\hfill
\begin{minipage}{0.45\textwidth}
  \centering
  \includegraphics[width=1.0\linewidth]{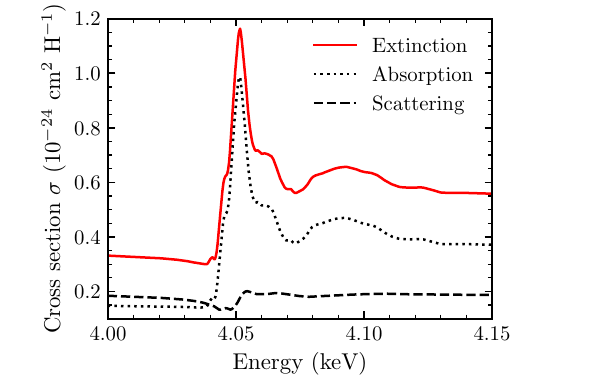}
  (b) 
\end{minipage}
\caption{(a) The calculated optical constants $n$ and $\kappa$ of calcium silicate. (b) Extinction (red solid), absorption (dash) and scattering (dotted line) corssection of calcium silicate.}
\label{fig:optical_constant_and_extiction_crosssection}
\end{figure}

To integrate the normalized XAS spectra with the astrophysical spectral fitting tool \texttt{SPEX}, the data were first interpolated onto the energy grid of the Henke tabulated cross-sections \citep{HENKE1993181}. The transmission cross-sections were obtained using the sample densities from Table \ref{tab:samples} and an estimated grain thickness of 5~$\mu$m to ensure the data remained within the optically thin regime. Subsequently, the spectra were rescaled by removing the slope in the Henke data to match the transmission edge intensities of the calcium photoabsorption feature, as defined by the Verner model in \texttt{SPEX}~\citep{1995A&AS..109..125V}. This process established the dust absorption cross-sections required for \texttt{SPEX} input parameters.
To convert the absorption coefficients were converted into extinction profiles, we calculated the complex refractive index, $m(\omega) = n(\omega) + ik(\omega)$. The imaginary component $k$ is derived directly from the absorption spectra, while the real part $n$ is calculated using the Kramers-Kronig relation~\citep{Watts:14}\footnote{\url{https://github.com/benajamin/kkcalc/blob/main/README.rst}}. An example of the result of the calculation of the complex refractive index of calcium silicate is shown in figure~\ref{fig:optical_constant_and_extiction_crosssection}(a).

In order to compute the extinction in the interstellar medium, we adopted a Mathis-Rumpl-Nordsieck (MRN) grain size distribution and applied the Anomalous Diffraction Theory (ADT) to account for scattering effects.
This data processing step follows the methodology established by~\citet{refId0Si,Costantini2022}. The complex optical constants and extinction cross-sections (figure~\ref{fig:optical_constant_and_extiction_crosssection}(b)) were subsequently exported for implementation in \texttt{SPEX} modeling.

\section{X-ray Source Selection and Spectral Modeling}\label{sec:X-ray_source_selection_and_spectral_modeling}
We selected three Low-Mass X-ray Binaries (LMXBs), GX340+00, GX13+1 and GX5-1, as background sources to probe the interstellar Ca K-edge. These targets were prioritized due to their high X-ray flux and significant interstellar extinction with reported column densities ($N_\mathrm{H}$) ranging from $3.1$ to $6.9\times 10^{22}$ $\mathrm{cm}^{-2}$~\citep{Costantini19,refId0Si, Rogantini18}. Source distances, coordinates, and corresponding OBSIDs are summarized in Table \ref{tab:source_list}.

Spectral analysis was performed using the \texttt{SPEX} fitting package. We adopted best fit parameters derived from archival \textit{Chandra} observations to simulate synthetic responses for \rm{XRISM}/\texttt{Resolve} and \rm{NewAthena}/X-IFU. These simulations evaluate the feasibility of resolving the XAFS features of calcium bearing dust analogs under the specific spectral resolutions of current and future missions. The detailed simulation parameters and results for each observatory are discussed in the following subsections.

\subsection{Fitting with Observational Chandra archive data in SPEX}

Observations were retrieved from the Chandra Transmission Grating Catalog (TGcat). Timed Exposure (TE) mode was prioritized over Continuous Clocking (CC) mode to investigate source properties at absorption edges. Observations in CC mode contain additional scattering within the CCD caused by the dust halos of LMXBs which may interfere with the shape of the Ca k-edge and produces shallow edge~\citep{2016ApJ...827...49S}\footnote{see also \url{https://cxc.harvard.edu/cal/Acis/Cal_prods/ccmode/ccmode_final_doc03.pdf}.} Although energies above 2 keV are less affected by the additional scattering~\citep{Kallman2019}, we cautiously selected the observations in TE mode to avoid any possible contamination in the edge.
We combined the +/- 1 orders of the HEG and +/- 1 orders of the MEG gratings using CIAO~\citep{2011AJ....141..129H, 2006SPIE.6270E..1VF}. However, the spectral fitting around the Ca K-edge covering 2 to 6 keV was restricted to HEG data, since HEG is less subject to pileup than MEG.

Spectral analysis was conducted using the SPEX software package, with initial parameters adopted from \citet{Zeegers18} and \citet{Rogantini18}. These parameters were constrained using archival \textit{Chandra} spectra to provide the physical inputs for our simulations. To account for the intervening interstellar medium (ISM) along the line of sight for all sources, we utilized the \texttt{hot} model with the temperature fixed at the lower limit of $1 \times 10^{-6}$ keV. At this temperature, the gas is effectively non-ionized, allowing the model to simulate pure photoelectric absorption from the cold interstellar foreground.

The individual sources were modeled using the following components:

\begin{itemize}
    \item GX 340+00: The continuum was modeled using a blackbody (\texttt{bb}) component for the thermal emission from the neutron star surface and a power-law (\texttt{pow}) to fit the high-energy non-thermal tail originating from the corona.
    
    \item GX 5-1: The continuum was modeled using a disk blackbody (\texttt{dbb}) for the thermal emission from the accretion disk and a Comptonization (\texttt{compt}) component to describe the hard tail produced by the scattering of seed photons in the corona.
    
    \item GX 13+1: The continuum was modeled using blackbody and Comptonization components to account for the thermal surface emission and the coronal tail, respectively. To represent the complex environment of this source, we included a photoionised absorption model (\texttt{xabs}) to account for the accretion disk wind. For a more rigorous treatment of the wind in this source, we refer the reader to \citet{2025ApJ...986...41R}, which utilizes the \texttt{pion} photoionization model.
\end{itemize}

\begin{table*}[ht]
    \centering
    \caption{Properties of the selected low-mass X-ray binaries (LMXBs). Distance and coordinate references are retrieved from \citet{2002A&A...391..923G}.}
    \label{tab:source_list}
    \begin{tabular}{llcccc}
        \hline
        Name & Chandra ObsId & Distance (kpc) & \multicolumn{2}{c}{Coordinates} & exposure time (ks) \\
        & & & $l$ (deg) & $b$ (deg) &  \\
        \hline
        GX 340+00 & 20099 & $11 \pm 0.3$ & 339.59 & $-0.08$ & $60.6$ \\
        GX 5-1    & 19449 & $9 \pm1$        & 5.08   & $-1.02$ & $80.5$\\
        GX 13+1   & 11816 & $7 \pm 1$    & 13.52  & $+0.11$ & $28.1$ \\
        \hline
    \end{tabular}
\end{table*}

\begin{table*}[htbp]  
\centering
\caption{Best fit parameters for GX 340+00,  GX 5-1, GX 13+1.}
\label{tab:fit_results}
\begin{tabular}{lccc}
\hline
Source & \multicolumn{1}{c}{\text{GX 340+00}} &  \multicolumn{1}{c}{\text{GX 5-1}} & \text{GX 13+1}  \\
Obsid &  20099 & 19449 & 11816  \\
\hline
$N_{\mathrm{H}}^{\mathrm{cold}} \,(10^{22}\,\mathrm{cm}^{-2})$ & $8.6 \pm 0.10$ & $5.20 \pm 0.03$ & $4.6 \pm 0.11$ \\ 
$k_BT_{\mathrm{bb}}$ (keV) &$ 1.40 \pm 0.02$ &$0.13 \pm 0.01$ & -\\
$\Gamma_{\mathrm{pow}} $ & $2.52 \pm 0.05$ & - & -  \\ 
$k_BT_{\mathrm{dbb}}$ (keV) & - & - & $ 0.34 \pm 0.01$  \\
$k_BT_{0,\mathrm{comt}}$ (keV) & - & $0.29 \pm 0.02$  & $0.44 \pm 0.02$ \\ 
$k_BT_{1,\mathrm{comt}}$ (keV) & - & $ 3.29 \pm 0.18$ &  $2.97^{+0.32}_{-1.10}$\\ 
$\tau_{\mathrm{compt}}$ & - & $5.09 \pm 0.14$ & $5.13\pm 0.21$\\ 
$N_{\mathrm{H}}^{\mathrm{xabs}}$ &- & - & $0.59 \pm 0.08$\\
$\log\xi^{\mathrm{xabs}}$& - & - & $ 5 \pm 0.28$ \\
$z\nu^{\mathrm{xabs}}$& - & - & $-571.84 \pm 56.61$ \\
$F_{0.5-2\,\mathrm{keV}}$ ($10^{-10}$ erg cm$^{-2}$ s$^{-1}$) &
$418.7 \pm 30.6$ &
$3.8 \pm 0.29$ &
$1.7 \pm 0.34$ \\
$F_{2-10\,\mathrm{keV}}$ ($10^{-9}$ erg cm$^{-2}$ s$^{-1}$) &
$9.397 \pm 0.69$ &
$18.0 \pm 0.08$ &
$6.9 \pm 0.99$\\
$C^2$/ $\nu_{new}$ & $3738/3237$ & $3752/3122$ & $3815/3781$ \\
\hline
\end{tabular}
\tablecomments{
The X-ray spectrum of GX 340+00 was modeled using a combination of blackbody and power law components, multiplied with the Hot model. Within the Hot model, the cold gas temperature is fixed at $T = 1 \times 10^{-6}\,\mathrm{K}$ to ensure a neutral stage; higher temperatures, such as $5 \times 10^{-4}\,\mathrm{K}$\ are insufficient to suppress ionized Ca II edge features~\citep{Costantini19}. Similarly, GX 5-1 were fitted using the Hot model with blackbody and comptonization components, with a disk blackbody specifically utilized for GX 13+1. The total fluxes for all sources were determined by the summation of the individual model components.}
\end{table*}

\begin{deluxetable*}{lllll}
\tabletypesize{\footnotesize}
\tablecaption{Performance comparison of X-ray instruments at 4.0 keV around the \ce{Ca} K-edge. \label{tab:instrument_comparison}}
\tablecolumns{5}
\tablewidth{0pt}
\tablehead{
\colhead{Instrument} & \colhead{Energy Range} & \colhead{Resolution} & \colhead{Effective Area} & \colhead{Reference} \\
\colhead{}           & \colhead{(keV)}        & \colhead{$\Delta E$ (eV)} & \colhead{(cm$^2$)}       & \colhead{}
}
\startdata
Chandra/ HETG   & 4--10   & $\sim$15\tablenotemark{a} & $\sim$20         & Chandra POG\tablenotemark{b} \\ 
\rm{XRISM}/\texttt{Resolve}  & 1.7--12 & $\sim$4.5  & 110--120           & \rm{XRISM} POG\tablenotemark{c}\\
\rm{NewAthena}/X-IFU & 0.2--12 & $\sim$3.5& ~3630  & \citet{2025ExA....59...18P} \\
\enddata
\tablenotetext{a}{Derived from table 8.1.\hfill}
\tablenotetext{b}{\url{https://cxc.harvard.edu/proposer/POG/html/}\hfill}
\tablenotetext{c}{\url{https://heasarc.gsfc.nasa.gov/docs/xrism/proposals/POG/}\hfill}
\end{deluxetable*}

\subsection{Results of Spectral Fitting and Constraints of Archival Data}
\label{Chandraarchive}

The best-fit parameters derived from the archival \textit{Chandra} HETG observations are summarized in Table~\ref{tab:fit_results}. The modeling yielded C-stat/dof ratios of 1.15, 1.2, and 1.0, with corresponding hydrogen column densities of $8.6 \times 10^{22}$~cm$^{-2}$, $5.2 \times 10^{22}$~cm$^{-2}$, and $4.6 \times 10^{22}$~cm$^{-2}$ for GX~340+00, GX~5-1, and GX~13+1, respectively. We adopt these values of the best-fit parameters as the physical basis for our simulation, utilizing the response matrices of the \rm{XRISM} and \rm{NewAthena} instruments.

\begin{figure}[h] \centering \includegraphics[width=1.1\linewidth]{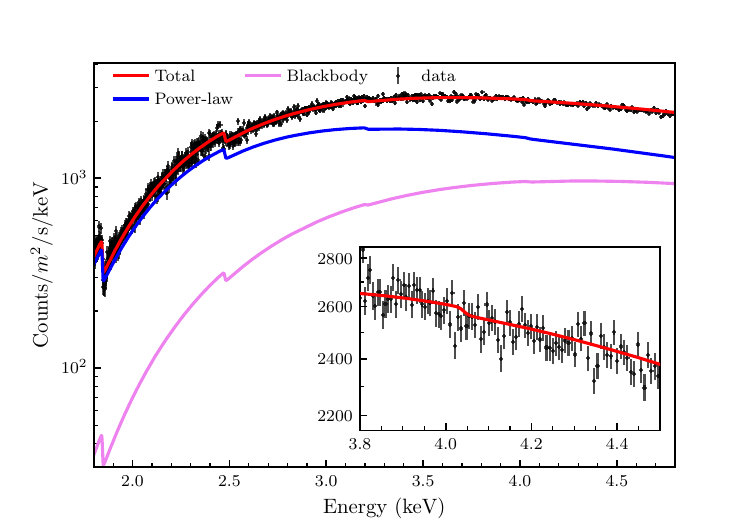} 
\caption{Best fit model for \textit{Chandra} archive data of GX~340+00 ($\mathrm{N_H}= 8.6 \times 10^{22} cm^{-2}$, obsid 20099). The spectrum was observed in TIMED (TE) mode for 60.6 ks and fitted with a combination of power law, blackbody model, hot and amol model in SPEX. The observation data and model are plotted without binning. The total model (red) is shown in the figure with individual contributions from the power-law (blue) and blackbody (magenta) models as a comparison.} 
\label{fig:fit_plot_models} \end{figure}

Figure~\ref{fig:fit_plot_models} illustrates the spectral fit for GX~340+00 (OBSID 20099), showing the relative contributions of the continuum components (\texttt{bb} and \texttt{pow} models) and the neutral ISM absorption (\texttt{hot} model). While the model can discern the edge and provide an estimate of the calcium abundance, the current energy resolution is not sufficient for a thorough investigation into the mineralogy or chemical composition of the dust. The resolution of \textit{Chandra} ($\sim$15~eV at 4.0~keV; see Table~\ref{tab:instrument_comparison}) is simply too coarse to resolve the intricate X-ray Absorption Fine Structure (XAFS) features required to distinguish between different calcium-bearing dust analogs.

In addition to the constraints on spectral resolution, the analysis is limited by the signal-to-noise ratio of the existing archival data. Currently, the \textit{Chandra} archive lacks observations of these sources exceeding 400~ks of exposure, resulting in insufficient photon statistics to differentiate subtle spectral structures from Poisson noise (refer to the inset in Figure~\ref{fig:fit_plot_models}). The combination of limited energy resolution and statistical sensitivity renders a definitive detection of solid-phase calcium dust absorption features unfeasible with \textit{Chandra} HETG. Consequently, we utilize simulations in the subsequent sections to assess the detection feasibility with next-generation instruments, specifically \rm{XRISM}-Resolve and \rm{NewAthena}-X-IFU.

\subsection{Simulations and results for \rm{XRISM} and \rm{NewAthena}}
\subsubsection{\rm{XRISM}}
The \rm{XRISM} observatory, currently in its operational phase, features the \texttt{Resolve} microcalorimeter which provides high resolution non-dispersive X-ray spectroscopy~\citep{2025PASJ...77S...1T}. Although the observatory is active, we utilize simulations to establish a feasibility baseline for detecting subtle mineralogical features under realistic observing conditions. Among the three sources selected for this work, GX 340+00 was utilized for the primary simulation due to its high column density ($N_H = 8.6 \times 10^{22}$~cm$^{-2}$), which provides the highest sensitivity to the Ca K-edge. Note that the response matrix we used for the preliminary investigation is with the closed gate valve case to present a realistic scenario. 

\begin{figure}[h]
\centering
\includegraphics[trim={0.5cm 0.5cm 0 0}, clip, width=1.0\columnwidth]{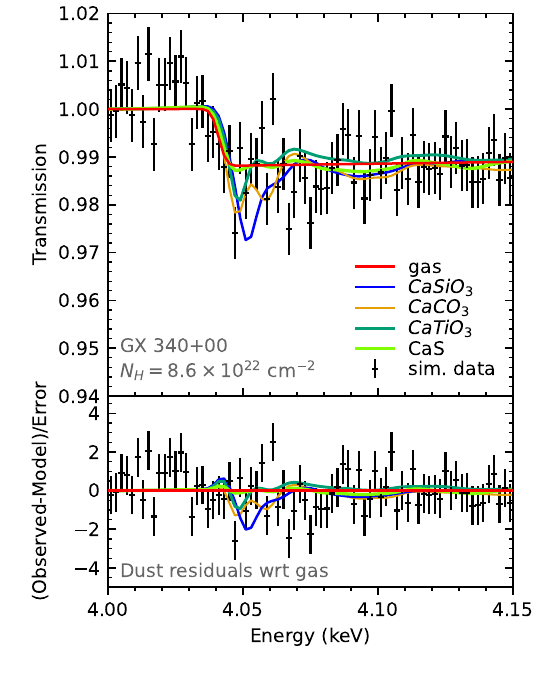}
\caption{Simulation for GX 340+00 with \rm{XRISM}-\texttt{Resolve} (gate valve closed.) Exposure time was set to 500ks and the data were binned by a factor of 4 for to enhance the ability to distinguish dust signatures from the simulated data (sim. data).}
\label{fig:XRISM_sim_results}
\end{figure}

To assess the detectability of the dust analogs from the simulation, we used a pure gas absorption edge as a reference to set a boundary for the spectrum in the absence of solid particles. To evaluate the dust absorption models, the hot model was set to zero calcium abundance, representing complete depletion from the gas phase. We have excluded the ionized gas (e.g., Ca II and Ca III) from our current simulation, as this study focuses specifically on the characterization of the model of the solid-phase dust grains. Furthermore, the individual dust in the amol models were convolved with the instrument response to evaluate the XAFS features of the dust analogs, as shown in figure~\ref{fig:XRISM_sim_results}. The data were binned by a factor of four to improve the signal to noise ratio. The top panels display the simulated spectra (labeled as 'sim. data') alongside the individual dust models in transmission, where the depth of the edge corresponds to the fraction of absorbed flux. Transmission (T(E)) is defined here as the ratio of the simulated observation ($\mathrm{F_{sim}(E)}$) to the continuum model $\mathrm{F_{cont}(E)}$ :

\begin{equation}
    T(E) = \frac{F_{\text{sim}}(E)}{F_{\text{cont}}(E)}
    \label{eq:transmission}
\end{equation}

This approach is consistent with \citet{2017A&A...599A.117Z} and \citet{Costantini19}; a larger transmission depth implies higher absorption.
The corresponding bottom panels show residuals computed in SPEX relative to the gas only baseline; these deviations fall within the $-2\%$ to $+1\%$ range, representing a significance below $1\sigma$ across the XAFS features (4.04–4.07~keV). Our simulations indicate that a 500~ks exposure, as documented in \citet{Costantini19}, sits at the detection threshold for discerning specific XAFS features of the dust analogs. While increased binning could improve the signal to noise ratio, it compromises the resolution needed to distinguish between dust species. Therefore, a multi-epoch approach using archival data is needed to increase exposure time and investigate the precise chemistry at the Ca K-edge.

\subsubsection{NewAthena}
For the simulations, we utilized the synthesized response matrix of the \rm{NewAthena} together with the best-fit parameters from \textit{Chandra} archive data in Section~\ref{Chandraarchive} as a baseline. The simulation time was set to 400ks in order to match that used by~\citet{Costantini19}. It is critical to note that the instrumental parameters used in~\citet{Costantini19} was based on the original Athena configuration with better energy resolution. The current effective spectral resolution can be found in Table~\ref{tab:instrument_comparison}.

Following the modeling procedure described in the previous section, the data were binned by a factor of three to improve the signal to noise. As illustrated in figure~\ref{fig:NewAthena_sim}, the top panels show the transmission models and simulated data, while the bottom panels display residuals relative to the gas only baseline. The high sensitivity of \rm{NewAthena/X-IFU} results in clearer deviations, even when the Ca K-edge absorption is subtle. Comparing the dust models to the gas-only baseline reveals XAFS features with a significance ranging from $-3\sigma$ to $+4\sigma$. This confirms that the features are clearly discernible above the noise floor at the sensitivity of \rm{NewAthena}/X-IFU. Furthermore, the results support the findings of \citet{Costantini19}, demonstrating that high-resolution spectroscopy can effectively distinguish between different calcium-bearing species, such as calcium carbonate ($\mathrm{CaCO_{3}}$) and calcium silicate ($\mathrm{CaSiO_{3}}$), under current and future mission configurations.

\section{Discussion}\label{sec:discussion}
\subsection{Instrumental Capabilities: \rm{XRISM} vs. \rm{NewAthena}}
While \rm{XRISM} currently serves as the benchmark for high resolution X-ray spectroscopy, \rm{NewAthena} is projected to offer higher observational efficiency due to its significantly larger effective area (see Table~\ref{tab:instrument_comparison}). This increased throughput facilitates the characterization of dust features with shorter exposure times compared to current \rm{XRISM} capabilities. By providing a more detailed characterization of the Ca K-edge environment, these new instruments enable a more precise determination of Ca abundance.

Despite these technological advances, the definitive identification of dominant mineral phases remains constrained by the limited samples of laboratory dust analogs currently available. Nevertheless, we provide a preliminary comparison of dust models based on their intrinsic XAFS features, as detailed in Section~\ref{sec:Spectral_and_Mineralogical_Identification}, establishing a foundation for future mineralogical studies with high resolution spectrometers.

Probing the Ca K-edge is inherently challenging due to its shallow transmission depth. As defined in Eq.~\ref{eq:transmission}, simulations with the three LMXBs in this study demonstrate edge depths ranging from 0.6\% to 1.2\% of the total flux. These weak transmission edge jumps stem from the relatively low interstellar abundance of calcium compared to more abundant elements such as Si, Mg, and Fe~\citep{2009LanB...4B..712L}. Furthermore, these features exhibit varying transmission depths across different lines of sight, reflecting the diverse physical and chemical properties of interstellar dust, including differences in grain size distribution and composition.

Consequently, detecting these features and accurately distinguishing between mineralogical phases necessitates the high signal-to-noise ratios achievable by \rm{XRISM} and \rm{NewAthena}. Beyond total transmission depth, the energy shift of the absorption edge relative to the gaseous Ca K-edge is a critical diagnostic, as it directly reveals the oxidation states and cation substitutions within the dust. To fully leverage the potential of these future missions, it is vital to expand the current dust analog database to account for these chemical variations, ensuring the necessary library exists to accurately interpret high-resolution data.

\begin{figure*}[htb!]
    \centering
    \begin{minipage}{0.32\textwidth}
        \centering
        \includegraphics[trim={0.1cm 0.5cm 0 0}, clip, width=\linewidth]{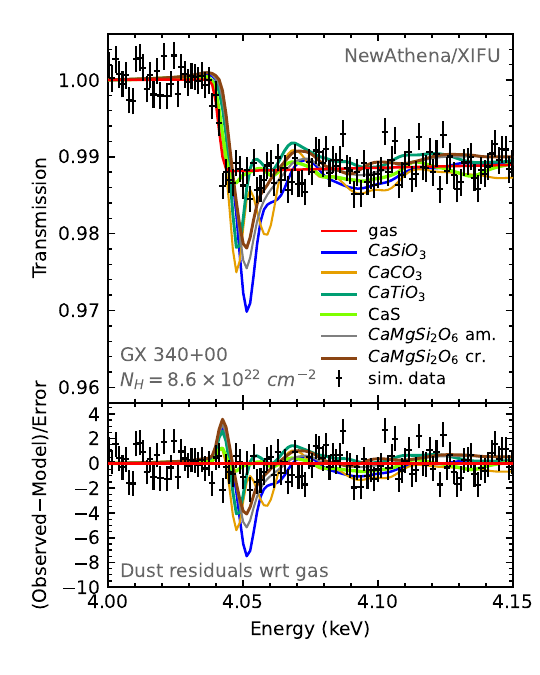}
        \small (a) GX 340+00 
    \end{minipage}\hfill
    \begin{minipage}{0.32\textwidth}
        \centering
        \includegraphics[trim={0.1cm 0.5cm 0 0}, clip, width=\linewidth]{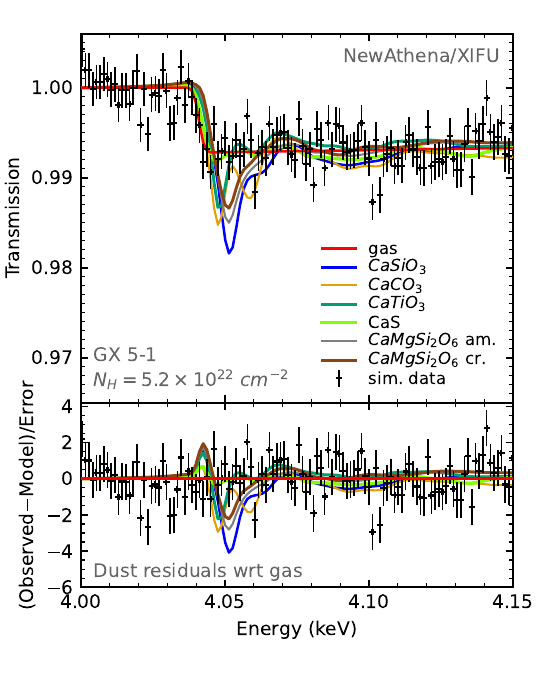}
        \small (b) GX 5-1
    \end{minipage}\hfill
    \begin{minipage}{0.32\textwidth}
        \centering
         \includegraphics[trim={0.1cm 0.5cm 0 0}, clip, width=\linewidth]{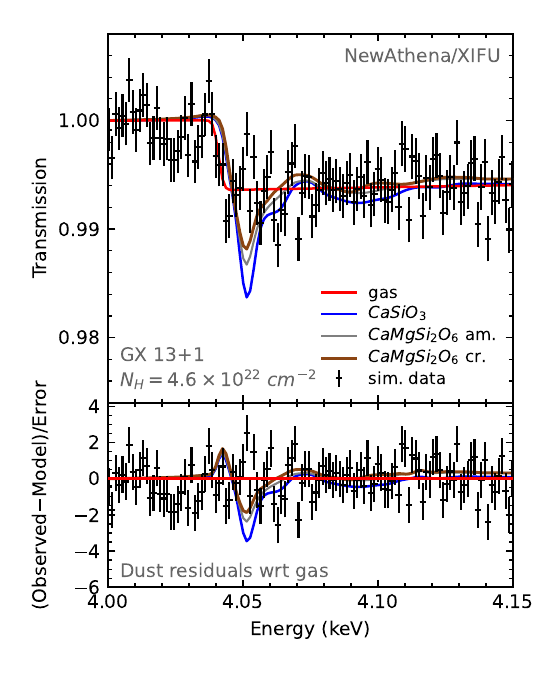}
        \small (c) GX 13+1
    \end{minipage}
    \caption{Simulated results with \rm{NewAthena/X-IFU} (400ks exposure, binning factor 3). Figure (a) and (b) plotted the dust model against simulated data (sim. data) along the line of sight from GX 340+00 and GX 5-1, respectively. Figure (c) displays the comparison of three silicate models in GX 13+1.}
    \label{fig:NewAthena_sim}
\end{figure*}

\subsection{Spectral and Mineralogical Identification}\label{sec:Spectral_and_Mineralogical_Identification}
Figure~\ref{fig:NewAthena_sim}(a) and (b) illustrates the spectral profiles of different dust groups for comparison with the \rm{NewAthena} response matrix, using GX340+00 and GX 5-1 as the background source, respectively. We simulate the dust model separately in order to characterize their XAFS features for the intervening dust along the line of sight. The $\mathrm{CaTiO_{3}}$ model exhibit a primary absorption peak at 4.0475 keV.This specific model likely overestimates the absorption intensity because the molecular structure assumes a one to one ratio of calcium to titanium. In reality, the solar abundance of titanium is approximately one order of magnitude lower than that of calcium, meaning such a high concentration of this specific mineral is unlikely in the interstellar medium~\citep{2009LanB...4B..712L}.

In contrast, $\mathrm{CaCO_{3}}$ displays a primary peak at a similar energy of 4.0468 keV. These species remain distinguishable because the $\mathrm{CaCO_{3}}$ spectrum produces a second prominent peak at 4.058 keV. While most dust models exhibit distinct profiles, the simulation reveals a degeneracy between the $\mathrm{CaS}$ model and purely gaseous absorption features. The $\mathrm{CaS}$ signal falls within the 1~$\sigma$ residual noise of the simulated gas only data, indicating that the noise effectively obscures the solid phase signatures of $\mathrm{CaS}$ at this sensitivity. Consequently, a purely gaseous edge profile in this energy range does not statistically exclude the presence of solid $\mathrm{CaS}$ grains. Since calcium is highly depleted and predominantly resides in the solid phase, $\mathrm{CaS}$ remains a viable candidate in these models. Resolving this ambiguity requires a dual approach: first, sulfur K-edge studies are necessary to provide independent verification of corresponding absorption; and second, significantly increased integration times are required to improve the signal to noise ratio beyond the current 3~$\sigma$ limitation. Current limitations in resolving these signatures are exacerbated by a binning factor of three, which reduces spectral resolution.

Figure~\ref{fig:NewAthena_sim}(c) presents the comparison between three silicate dusts: calcium silicate ($\mathrm{CaSiO_{3}}$), and crystalline and amorphous diopside ($\mathrm{CaMgSi_{2}O_{6}}$). Our simulations reveal that the spectral features of Ca-bearing silicate dust analogs which are governed by the interplay of stoichiometry, lattice order, and local coordination geometry. Specifically, the maximum absorption intensity scales directly with the calcium content within the mineral stoichiometry; for example, calcium silicate ($\mathrm{CaSiO_{3}}$) contains a higher proportion of $\mathrm{Ca^{2+}}$ relative to the total cations in its stoichiometric composition than Ca–Mg silicates such as diopside ($\mathrm{CaMgSi_2O_6}$). In addition, from the comparison between crystalline and amorphous diopside, we can see that the amorphous crystal structure presents a stronger first peak. Overall, they all share a similar crystal structure, which implies similar Ca-O binding in the local environment. Although it is challenging to discern between the silicate grains, the feature at 4.04 keV can certainly be an indicator of their existence. The prominence of these absorption features is modulated by the  column density along the line of sight and the intrinsic flux of the background X-ray source, as illustrated in figure~\ref{fig:NewAthena_sim}(c).  Furthermore, multi-edge fitting will likely resolve current degeneracies associated with crystalline silicates identifications.

\subsection{Future expansion of the dust analogs model and integration in IR}
Beyond the analogs examined in this work, expanding the AMOL database to include a broader  collection of Ca-bearing dust analog models is essential for achieving a more comprehensive investigation of theCa K-edge. Primary candidates for inclusion include Ca-Al rich phases, such as hibonite ($\mathrm{CaAl_{12}O_{19}}$), and members of the melilite group solid solutions (e.g., akermanite ($\mathrm{Ca_2MgSi_2O_7}$) and gehlenite ($\mathrm{Ca_2Al_2Si_2O_7}$).

Incorporation of specific pyroxene group minerals is similarly warranted. Due to the nature of silicate crystal chemistry, the large ionic radius of $\mathrm{Ca^{2+}}$ typically limits its significant substitution into olivine or orthopyroxene lattices~\citep{WOS:A1981LC75800001, 2017GeCoA.219...44W}. Consequently, calcium is primarily accommodated within clinopyroxenes such as diopside, augite and hedenbergite. Beyond these silicate phases, dolomite can also be studied as a variation of the carbonate minerals. Finally, investigating solid solution series by accounting for varying stoichiometric metal ratios and degrees of crystallinity is essential for a more precise characterization of the interstellar environment.  

Another approach is through multi-wavelength studies and incorporate IR spectroscopy, which requires the identification of LMXB counterparts. However, this approach is currently constrained by the fact that only GX 13+1 has a confirmed IR detection among the three sources mentioned in this study~\citep{1999MNRAS.306..417B}. Furthermore, IR detection is challenged by the weak intrinsic intensity of the calcium vibrational modes. Nevertheless, quantum chemical simulations of calcium silicate hydrate clusters suggest a diagnostic vibrational feature near $14~\mu\text{m}$ \citep{Bilalbegovic14}. Utilizing these theoretical fingerprints to guide future coordinated JWST and X-ray observation will be vital for bridging the gap between gas phase depletion and solid state mineralogy.

\section{Conclusion}\label{sec:summary}
In this study, we provide an extensive set of calcium dust extinction models based on the laboratory measurement of seven calcium-bearing dust analogs. These measurements were conducted at the National Synchrotron Radiation Research Center (NSRRC) in Taiwan using the TPS32A and TLS16A1 beamlines. These models extend the existing \texttt{AMOL} library for the calcium bearing dust analogs at the calcium K-edge and will be made available for future updates of the \texttt{SPEX} spectral fitting package.

The primary conclusions of this work are as follow:
\begin{itemize}
    \item The Ca K-edge can be resolved using the \rm{XRISM} and \rm{NewAthena} response calorimeters. Simulations demonstrate that the spectral resolution of these instruments is sufficient to identify distinct XAFS signatures, allowing for the characterization of calcium dust species in the ISM. 
    
    \item Calcium carbonates and calcium silicates are spectrally distinct at the Ca K-edge. Differences in both first peak energies and XAFS features result in high residuals in our simulations. This allows for a clear determination of the dominant mineralogical host of interstellar calcium, validating and extending the predictions in~\citet{Costantini19}.
    
    \item Solid phase $\mathrm{CaS}$ is spectrally degenerate with gaseous calcium at Ca K-edge. Because the residuals for CaS are low in our simulations, the Ca K-edge alone does not uniquely constrain the presence of sulfide grains. Consequently, multi edge analysis including the sulfur K-edge is essential to resolve this degeneracy and characterize complex chemical mixtures in the X-ray regime. 
\end{itemize}

We look forward to the powerful observations from next generation X-ray observatories. While early public data from \rm{XRISM} is becoming available, longer exposure times will be required to match the sensitivity of our simulations and fully test these models against real scenarios. Both missions provide transformative capabilities for the investigation of calcium dust in the interstellar medium.

\section{Acknowledgements}

The authors acknowledge support from the Institute of Astronomy and Astrophysics, Academia Sinica (ASIAA), and the National Science and Technology Council (NSTC) in Taiwan through grants 112-2112-M-001-030-, 113-2112-M-001-008-, and 114-2112-M-001-001-. 
We thank the experimental facility and the technical services provided by the National Synchrotron Radiation Research Center (NSRRC), a national user facility supported by the National Science and Technology Council of Taiwan (NSTC), Taiwan (R.O.C.). The data used in this work were collected at beamlines TLS16A1 and TPS32A, with time allocations under 2023-2-041-1 and 2023-2-041-2 (PI: H. Shang) at NSRRC, Hsinchu, Taiwan (R.O.C.). This paper employs a list of Chandra datasets, obtained by the Chandra X-ray Observatory, contained in the Chandra Data Collection (CDC) 659 ~\href{https://doi.org/10.25574/cdc.659}{doi:10.25574/cdc.659}. This research has made use of data obtained from The Chandra Grating-Data Archive and Catalog (TGCat) \citet{2011AJ....141..129H}, and CIAO version 4.17 provided by the Chandra X-ray Center (CXC).
We thank the experimental facility and the technical services provided by the National Synchrotron Radiation Research Center (NSRRC), a national user facility supported by the National Science and Technology Council of Taiwan (NSTC), Taiwan (R.O.C.). The data used in this work were collected at beamlines TLS16A1 and TPS32A, with time allocations under 2023-2-041-1 and 2023-2-041-2 (PI: H. Shang) at NSRRC, Hsinchu, Taiwan (R.O.C.). 
We thank Gabriele Born and Birgit Begemann for producing the diopside samples in the Jena Laboratory, and Susanne Goepel for providing these samples, as well as the additional calcium silicate. This manuscript made use of Google Gemini (version 3.1 Flash-lite) for grammar and spelling corrections and to refine the text for improved clarity and flow. The authors have critically reviewed and edited the final output to ensure accuracy and alignment with the original research findings.

\software{In addition to the software mentioned in the text, this work has made use of the following modules: Matplotlib~\citep{Hunter:2007}, Numpy~\citep{harris2020array}, and the IDL Astronomy User's Library,} available online via \url{https://github.com/wlandsman/IDLAstro#idlastro.}

\bibliography{sample631}{}

@ARTICLE{1994ApJ...424..748C,
       author = {{Crinklaw}, Greg and {Federman}, S.~R. and {Joseph}, Charles L.},
        title = "{The Depletion of Calcium in the Interstellar Medium}",
      journal = {\apj},
         year = 1994,
        month = apr,
       volume = {424},
        pages = {748},
          doi = {10.1086/173927},
       adsurl = {https://ui.adsabs.harvard.edu/abs/1994ApJ...424..748C}
}

@ARTICLE{Costantini19,
       author = {{Costantini}, E. and {Zeegers}, S.~T. and {Rogantini}, D. and {de Vries}, C.~P. and {Tielens}, A.~G.~G.~M. and {Waters}, L.~B.~F.~M.},
        title = "{X-ray extinction from interstellar dust. Prospects of observing carbon, sulfur, and other trace elements}",
      journal = {\aap},
         year = 2019,
        month = sep,
       volume = {629},
          eid = {A78},
        pages = {A78},
          doi = {10.1051/0004-6361/201833820},
archivePrefix = {arXiv},
       eprint = {1906.08653},
 primaryClass = {astro-ph.GA},
       adsurl = {https://ui.adsabs.harvard.edu/abs/2019A&A...629A..78C}
}

@article{Bilalbegovic14,
    author = {Bilalbegović, G. and Maksimović, A. and Mohaček-Grošev, V.},
    title = "{Do cement nanoparticles exist in space?}",
    journal = {Monthly Notices of the Royal Astronomical Society},
    volume = {442},
    number = {2},
    pages = {1319-1325},
    year = {2014},
    month = {06},
    issn = {0035-8711},
    doi = {10.1093/mnras/stu869},
    url = {https://doi.org/10.1093/mnras/stu869},
    eprint = {https://academic.oup.com/mnras/article-pdf/442/2/1319/5699785/stu869.pdf},
}

@ARTICLE{Ferrarotti05,
       author = {{Ferrarotti}, A.~S. and {Gail}, H. -P.},
        title = "{Mineral formation in stellar winds. V. Formation of calcium carbonate}",
      journal = {\aap},
         year = 2005,
        month = feb,
       volume = {430},
        pages = {959-965},
          doi = {10.1051/0004-6361:20041856},
       adsurl = {https://ui.adsabs.harvard.edu/abs/2005A&A...430..959F}
}

@ARTICLE{Kempernat02,
       author = {{Kemper}, F. and {J{\"a}ger}, C. and {Waters}, L.~B.~F.~M. and {Henning}, Th. and {Molster}, F.~J. and {Barlow}, M.~J. and {Lim}, T. and {de Koter}, A.},
        title = "{Detection of carbonates in dust shells around evolved stars}",
      journal = {\nat},
         year = 2002,
        month = jan,
       volume = {415},
       number = {6869},
        pages = {295-297},
          doi = {10.1038/415295a},
       adsurl = {https://ui.adsabs.harvard.edu/abs/2002Natur.415..295K}
}

@article{HENKE1993181,
title = {X-Ray Interactions: Photoabsorption, Scattering, Transmission, and Reflection at E = 50-30,000 eV, Z = 1-92},
journal = {Atomic Data and Nuclear Data Tables},
volume = {54},
number = {2},
pages = {181-342},
year = {1993},
issn = {0092-640X},
doi = {https://doi.org/10.1006/adnd.1993.1013},
url = {https://www.sciencedirect.com/science/article/pii/S0092640X83710132},
author = {B.L. Henke and E.M. Gullikson and J.C. Davis}
}

@ARTICLE{2011Jones,
       author = {{Jones}, A.~P. and {Nuth}, J.~A.},
        title = "{Dust destruction in the ISM: a re-evaluation of dust lifetimes}",
      journal = {\aap},
         year = 2011,
        month = jun,
       volume = {530},
          eid = {A44},
        pages = {A44},
          doi = {10.1051/0004-6361/201014440},
       adsurl = {https://ui.adsabs.harvard.edu/abs/2011A&A...530A..44J}
}

@BOOK{Draine11,
       author = {{Draine}, Bruce T.},
        title = "{Physics of the Interstellar and Intergalactic Medium}",
         year = 2011,
       adsurl = {https://ui.adsabs.harvard.edu/abs/2011piim.book.....D}
}

@ARTICLE{Sanghani22,
       author = {{Sanghani}, Manish N. and {Lajaunie}, Luc and {Marhas}, Kuljeet Kaur and {Rickard}, William D.~A. and {Hsiao}, Silver Sung-Yun and {Peeters}, Zan and {Shang}, Hsien and {Lee}, Der-Chuen and {Calvino}, Jos{\'e}. J. and {Bizzarro}, Martin},
        title = "{Microstructural and Chemical Investigations of Presolar Silicates from Diverse Stellar Environments}",
      journal = {\apj},
         year = 2022,
        month = feb,
       volume = {925},
       number = {2},
          eid = {110},
        pages = {110},
          doi = {10.3847/1538-4357/ac3332},
       adsurl = {https://ui.adsabs.harvard.edu/abs/2022ApJ...925..110S}
}

@ARTICLE{Rogantini18,
       author = {{Rogantini}, D. and {Costantini}, E. and {Zeegers}, S.~T. and {de Vries}, C.~P. and {Bras}, W. and {de Groot}, F. and {Mutschke}, H. and {Waters}, L.~B.~F.~M.},
        title = "{Investigating the interstellar dust through the Fe K-edge}",
      journal = {\aap},
         year = 2018,
        month = jan,
       volume = {609},
          eid = {A22},
        pages = {A22},
          doi = {10.1051/0004-6361/201731664},
archivePrefix = {arXiv},
       eprint = {1709.05359},
 primaryClass = {astro-ph.HE},
       adsurl = {https://ui.adsabs.harvard.edu/abs/2018A&A...609A..22R}
}

@INPROCEEDINGS{kaastra1996_spex,
   author = {{Kaastra}, J.~S. and {Mewe}, R. and {Nieuwenhuijzen}, H.},
    title = "{SPEX: a new code for spectral analysis of X {\amp} UV spectra.}",
booktitle = {UV and X-ray Spectroscopy of Astrophysical and Laboratory Plasmas},
     year = 1996,
   editor = {{Yamashita}, K. and {Watanabe}, T.},
    pages = {411-414},
   adsurl = {http://adsabs.harvard.edu/abs/1996uxsa.conf..411K}
}

@article{article,
author = {Newville, Matthew},
year = {2004},
month = {01},
pages = {},
title = {Fundamentals of XAFS},
volume = {78},
journal = {Consortium for Advanced Radiation Sources, University of Chicago (USA)[http://xafs. org]},
doi = {10.2138/rmg.2014.78.2}
}

@article{PhysRevB.46.3283,
  title = {Full correction of the self-absorption in soft-fluorescence extended x-ray-absorption fine structure},
  author = {Tr\"oger, L. and Arvanitis, D. and Baberschke, K. and Michaelis, H. and Grimm, U. and Zschech, E.},
  journal = {Phys. Rev. B},
  volume = {46},
  issue = {6},
  pages = {3283--3289},
  numpages = {0},
  year = {1992},
  month = {Aug},
  publisher = {American Physical Society},
  doi = {10.1103/PhysRevB.46.3283},
  url = {https://link.aps.org/doi/10.1103/PhysRevB.46.3283}
}

@PHDTHESIS{Zeegers18,
       author = {{Zeegers}, Sascha Tamara},
        title = "{X-ray spectroscopy of interstellar dust: From the laboratory to the Galaxy}",
       school = {Leiden University, Netherlands},
         year = 2018,
        month = jan,
       adsurl = {https://ui.adsabs.harvard.edu/abs/2018PhDT........73Z}
}

@ARTICLE{2025ExA....59...18P,
       author = {{Peille}, Philippe and {Barret}, Didier and {Cucchetti}, Edoardo and {Albouys}, Vincent and {Piro}, Luigi and {Simionescu}, Aurora and {Cappi}, Massimo and {Bellouard}, Elise and {C{\'e}nac-Morth{\'e}}, C{\'e}line and {Daniel}, Christophe and {Pradines}, Alice and {Finoguenov}, Alexis and {Kelley}, Richard and {Mas-Hesse}, J. Miguel and {Paltani}, St{\'e}phane and {Rauw}, Gregor and {Rozanska}, Agata and {Svoboda}, Jiri and {Wilms}, Joern and {Audard}, Marc and {Bozzo}, Enrico and {Costantini}, Elisa and {Dadina}, Mauro and {Dauser}, Thomas and {Decourchelle}, Anne and {den Herder}, Jan-Willem and {Goldwurm}, Andrea and {Jonker}, Peter and {Markowitz}, Alex and {Mendez}, Mariano and {Miniutti}, Giovanni and {Molendi}, Silvano and {Nicastro}, Fabrizio and {Pajot}, Fran{\c{c}}ois and {Pointecouteau}, Etienne and {Pratt}, Gabriel W. and {Schaye}, Joop and {Vink}, Jacco and {Webb}, Natalie and {Bandler}, Simon and {Barbera}, Marco and {Ceballos}, Maria Teresa and {Charles}, Ivan and {den Hartog}, Roland and {Doriese}, W. Bertrand and {Duval}, Jean-Marc and {Gatti}, Flavio and {Jackson}, Brian and {Kilbourne}, Caroline and {Macculi}, Claudio and {Martin}, Sylvain and {Parot}, Yann and {Porter}, Frederick and {Pr{\^e}le}, Damien and {Ravera}, Laurent and {Smith}, Stephen and {Soucek}, Jan and {Thibert}, Tanguy and {Tuominen}, Eija and {Acero}, Fabio and {Ettori}, Stefano and {Grosso}, Nicolas and {Kaastra}, Jelle and {Mazzotta}, Pasquale and {Miller}, Jon and {Sciortino}, Salvatore and {Beaumont}, Sophie and {D'Andrea}, Matteo and {de Plaa}, Jelle and {Eckart}, Megan and {Gottardi}, Luciano and {Leutenegger}, Maurice and {Lotti}, Simone and {Molin}, Alexei and {Natalucci}, Lorenzo and {Adil}, Muhammad Qazi and {Argan}, Andrea and {Cavazzuti}, Elisabetta and {Fiorini}, Mauro and {Khosropanah}, Pourya and {Medinaceli Villegas}, Eduardo and {Minervini}, Gabriele and {Perry}, James and {Pinsard}, Frederic and {Raulin}, Desi and {Rigano}, Manuela and {Roelfsema}, Peter and {Schwander}, Denis and {Terron}, Santiago and {Torrioli}, Guido and {Ullom}, Joel and {Zuchniak}, Monika and {Chaoul}, Laurence and {Torrejon}, Jose Miguel and {Brachet}, Frank and {Cobo}, Beatriz and {Durkin}, Malcolm and {Fioretti}, Valentina and {Geoffray}, Herv{\'e} and {Jacques}, Lionel and {Kirsch}, Christian and {Lo Cicero}, Ugo and {Adams}, Joseph and {Gloaguen}, Emilie and {Gonzalez}, Manuel and {Hull}, Samuel and {Jellyman}, Erik and {Kiviranta}, Mikko and {Sakai}, Kazuhiro and {Taralli}, Emanuele and {Vaccaro}, Davide and {van der Hulst}, Paul and {van der Kuur}, Jan and {van Leeuwen}, Bert-Joost and {van Loon}, Dennis and {Wakeham}, Nicholas and {Auricchio}, Natalia and {Brienza}, Daniele and {Cheatom}, Oscar and {Franssen}, Philippe and {Julien}, Sabine and {Le Mer}, Isabelle and {Moirin}, David and {Silva}, Vitor and {Todaro}, Michela and {Clerc}, Nicolas and {Coleiro}, Alexis and {Ptak}, Andy and {Puccetti}, Simonetta and {Surace}, Christian and {Abdoelkariem}, Shariefa and {Adami}, Christophe and {Aicardi}, Corinne and {Andr{\'e}}, J{\'e}r{\^o}me and {Angelinelli}, Matteo and {Anvar}, Shebli and {Arnaldi}, Luis Horacio and {Attard}, Anthony and {Audley}, Damian and {Bancel}, Florian and {Banks}, Kimberly and {Bernard}, Vivian and {Bij de Vaate}, Jan Geralt and {Bonino}, Donata and {Bonnamy}, Anthony and {Bonny}, Patrick and {Boreux}, Charles and {Bounab}, Ayoub and {Brigitte}, Ma{\"\i}mouna and {Bruijn}, Marcel and {Brysbaert}, Cl{\'e}ment and {Bulgarelli}, Andrea and {Calarco}, Simona and {Camus}, Thierry and {Canourgues}, Florent and {Capobianco}, Vito and {Cardiel}, Nicolas and {Celasco}, Edvige and {Chen}, Si and {Chervenak}, James and {Chiarello}, Fabio and {Clamagirand}, S{\'e}bastien and {Coeur-Joly}, Odile and {Corcione}, Leonardo and {Coriat}, Mickael and {Coulet}, Anais and {Courty}, Bernard and {Coynel}, Alexandre and {D'Ai}, Antonino and {Dambrauskas}, Eugenio and {D'anca}, Fabio and {Dauner}, Lea and {De Gerone}, Matteo and {DeNigris}, Natalie and {Dercksen}, Johannes and {de Wit}, Martin and {Dieleman}, Pieter and {DiPirro}, Michael and {Doumayrou}, Eric and {Duband}, Lionel and {Dubbeldam}, Luc and {Dupieux}, Michel and {Dupourqu{\'e}}, Simon and {Durand}, Jean Louis and {Eckert}, Dominique and {Ferrando}, Philippe and {Ferrari Barusso}, Lorenzo and {Finkbeiner}, Fred and {Fiocchi}, Mariateresa and {Fossecave}, Herv{\'e} and {Gabici}, Stefano and {Gallucci}, Giovanni and {Gant}, Florent and {Gao}, Jian-Rong and {Gastaldello}, Fabio and {Genolet}, Ludovic and {Ghizzardi}, Simona and {Giovannini}, Elisa and {Giustini}, Margherita and {Givaudan}, Alain},
        title = "{The X-ray Integral Field Unit at the end of the Athena reformulation phase}",
      journal = {Experimental Astronomy},
         year = 2025,
        month = apr,
       volume = {59},
       number = {2},
          eid = {18},
        pages = {18},
          doi = {10.1007/s10686-025-09984-w},
archivePrefix = {arXiv},
       eprint = {2502.10866},
 primaryClass = {astro-ph.IM},
       adsurl = {https://ui.adsabs.harvard.edu/abs/2025ExA....59...18P}
}

@article{Liang_2025,
doi = {10.1088/1742-6596/3010/1/012121},
url = {https://doi.org/10.1088/1742-6596/3010/1/012121},
year = {2025},
month = {may},
publisher = {IOP Publishing},
volume = {3010},
number = {1},
pages = {012121},
author = {Liang, Cheng-Chih and Tu, Jia-Lin and Lee, Ming-Han and Chen, Hsin-Wei and Chiang, Liang-Chih and Lee, Chang-Sheng and Chan, Chi and Chang, Chia-Feng and Chang, Chao-Yu and Yu, Kuan-Li and Liu, Din-Goa and Huang, Yu-Cheng and Peng, Po-Yang and Lu, Ying-Jui and Chen, Chi-Liang},
title = {Data Acquisition Software for the Tender X-ray Absorption Spectroscopy Beamline at the Taiwan Photon Source},
journal = {Journal of Physics: Conference Series}
}

@article{ refId0Si,
	author = {{Zeegers, S. T.} and {Costantini, E.} and {Rogantini, D.} and {de Vries, C. P.} and {Mutschke, H.} and {Mohr, P.} and {de Groot, F.} and {Tielens, A. G. G. M.}},
	title = {Dust absorption and scattering in the silicon K-edge},
	DOI= "10.1051/0004-6361/201935050",
	url= "https://doi.org/10.1051/0004-6361/201935050",
	journal = {A\&A},
	year = 2019,
	volume = 627,
	pages = "A16",
}

@article{Ravel:ph5155,
author = "Ravel, B. and Newville, M.",
title = "{{\it ATHENA}, {\it ARTEMIS}, {\it HEPHAESTUS}: data analysis for X-ray absorption spectroscopy using {\it IFEFFIT}}",
journal = "Journal of Synchrotron Radiation",
year = "2005",
volume = "12",
number = "4",
pages = "537--541",
month = "Jul",
doi = {10.1107/S0909049505012719},
url = {https://doi.org/10.1107/S0909049505012719},
}

@ARTICLE{2022arXiv220205399X,
       author = {{XRISM Science Team}},
        title = "{XRISM Quick Reference}",
      journal = {arXiv e-prints},
         year = 2022,
        month = feb,
          eid = {arXiv:2202.05399},
        pages = {arXiv:2202.05399},
          doi = {10.48550/arXiv.2202.05399},
archivePrefix = {arXiv},
       eprint = {2202.05399},
 primaryClass = {astro-ph.IM},
       adsurl = {https://ui.adsabs.harvard.edu/abs/2022arXiv220205399X}
}

@ARTICLE{2002A&A...391..923G,
       author = {{Grimm}, H.-J. and {Gilfanov}, M. and {Sunyaev}, R.},
        title = "{The Milky Way in X-rays for an outside observer. Log(N)-Log(S) and luminosity function of X-ray binaries from RXTE/ASM data}",
      journal = {\aap},
         year = 2002,
        month = sep,
       volume = {391},
        pages = {923-944},
          doi = {10.1051/0004-6361:20020826},
archivePrefix = {arXiv},
       eprint = {astro-ph/0109239},
 primaryClass = {astro-ph},
       adsurl = {https://ui.adsabs.harvard.edu/abs/2002A&A...391..923G}
}

@ARTICLE{2009LanB...4B..712L,
       author = {{Lodders}, K. and {Palme}, H. and {Gail}, H.-P.},
        title = "{Abundances of the Elements in the Solar System}",
      journal = {Landolt B{\"o}rnstein},
         year = 2009,
        month = jan,
       volume = {4B},
        pages = {712},
          doi = {10.1007/978-3-540-88055-4_34},
archivePrefix = {arXiv},
       eprint = {0901.1149},
 primaryClass = {astro-ph.EP},
       adsurl = {https://ui.adsabs.harvard.edu/abs/2009LanB...4B..712L}
}

@ARTICLE{2025NatAs...9.1356D,
       author = {{Das}, Priyam and {Seitenzahl}, Ivo R. and {Ruiter}, Ashley J. and {R{\"o}pke}, Friedrich K. and {Pakmor}, R{\"u}diger and {Vogt}, Fr{\'e}d{\'e}ric P.~A. and {Collins}, Christine E. and {Ghavamian}, Parviz and {Sim}, Stuart A. and {Williams}, Brian J. and {Taubenberger}, Stefan and {Laming}, J. Martin and {Suherli}, Janette and {Sutherland}, Ralph and {Rodr{\'\i}guez-Segovia}, Nicol{\'a}s},
        title = "{Calcium in a supernova remnant as a fingerprint of a sub-Chandrasekhar-mass explosion}",
      journal = {Nature Astronomy},
         year = 2025,
        month = sep,
       volume = {9},
        pages = {1356-1365},
          doi = {10.1038/s41550-025-02589-5},
       adsurl = {https://ui.adsabs.harvard.edu/abs/2025NatAs...9.1356D}
}

@article{Watts:14,
author = {Benjamin Watts},
journal = {Opt. Express},
number = {19},
pages = {23628--23639},
publisher = {Optica Publishing Group},
title = {Calculation of the Kramers-Kronig transform of X-ray spectra by a piecewise Laurent polynomial method},
volume = {22},
month = {Sep},
year = {2014},
url = {https://opg.optica.org/oe/abstract.cfm?URI=oe-22-19-23628},
doi = {10.1364/OE.22.023628},
}

@software{2024zndo..10822753K,
       author = {{Kaastra}, J.~S. and {Raassen}, A.~J.~J. and {de Plaa}, J. and {Gu}, Liyi},
        title = "{SPEX X-ray spectral fitting package}",
         year = 2024,
        month = mar,
          eid = {10.5281/zenodo.10822753},
          doi = {10.5281/zenodo.10822753},
      version = {3.08.00},
    publisher = {Zenodo},
       adsurl = {https://ui.adsabs.harvard.edu/abs/2024zndo..10822753K}
}

@ARTICLE{2022FrASS...9.8217T,
       author = {{Tielens}, A.~G.~G.~M.},
        title = "{Dust Formation in Astrophysical Environments: The Importance of Kinetics}",
      journal = {Frontiers in Astronomy and Space Sciences},
         year = 2022,
        month = may,
       volume = {9},
          eid = {908217},
        pages = {908217},
          doi = {10.3389/fspas.2022.908217},
archivePrefix = {arXiv},
       eprint = {2206.01548},
 primaryClass = {astro-ph.EP},
       adsurl = {https://ui.adsabs.harvard.edu/abs/2022FrASS...9.8217T}
}

@INPROCEEDINGS{2022JPhCS2380a2041L,
       author = {{Liu}, D.~G. and {Lee}, M.~H. and {Lu}, Y.~J. and {Liang}, C.~C. and {Chang}, C.~F. and {Tu}, J.~L. and {Lee}, J.~F. and {Chen}, C.~L.},
        title = "{Design of Tender X-ray Absorption Spectroscopy Beamline in Taiwan Photon Source}",
    booktitle = {Journal of Physics Conference Series},
         year = 2022,
       series = {Journal of Physics Conference Series},
       volume = {2380},
        month = dec,
    publisher = {IOP},
          eid = {012041},
        pages = {012041},
          doi = {10.1088/1742-6596/2380/1/012041},
       adsurl = {https://ui.adsabs.harvard.edu/abs/2022JPhCS2380a2041L}
}

@book{thompson2009x,
  title={X-ray Data Booklet},
  author={Thompson, A.C.},
  url={https://xdb.lbl.gov/},
  year={2009},
  publisher={Lawrence Berkeley National Laboratory, University of California}
}

@ARTICLE{2024Univ...10..148B,
       author = {{Boccioli}, Luca and {Roberti}, Lorenzo},
        title = "{The Physics of Core-Collapse Supernovae: Explosion Mechanism and Explosive Nucleosynthesis}",
      journal = {Universe},
         year = 2024,
        month = mar,
       volume = {10},
       number = {3},
          eid = {148},
        pages = {148},
          doi = {10.3390/universe10030148},
archivePrefix = {arXiv},
       eprint = {2403.12942},
 primaryClass = {astro-ph.SR},
       adsurl = {https://ui.adsabs.harvard.edu/abs/2024Univ...10..148B}
}

@ARTICLE{2025PASJ...77S.107C,
       author = {{Corrales}, L{\'\i}a and {Costantini}, Elisa and {Zeegers}, Sascha and {Gu}, Liyi and {Takahashi}, Hiromitsu and {Moutard}, David and {Shidatsu}, Megumi and {Miller}, Jon M. and {Mizumoto}, Misaki and {Smith}, Randall K. and {Ballhausen}, Ralf and {Chakraborty}, Priyanka and {D{\'\i}az Trigo}, Mar{\'\i}a and {Ludlam}, Renee and {Nakagawa}, Takao and {Psaradaki}, Ioanna and {Yamada}, Shinya and {Kilbourne}, Caroline A.},
        title = "{XRISM insights for interstellar sulfur}",
      journal = {\pasj},
         year = 2025,
        month = jun,
       volume = {77},
          eid = {psaf068},
        pages = {psaf068},
          doi = {10.1093/pasj/psaf068},
archivePrefix = {arXiv},
       eprint = {2506.08751},
 primaryClass = {astro-ph.GA},
       adsurl = {https://ui.adsabs.harvard.edu/abs/2025PASJ...77S.107C}
}

@ARTICLE{1999MNRAS.306..417B,
       author = {{Bandyopadhyay}, R.~M. and {Shahbaz}, T. and {Charles}, P.~A. and {Naylor}, T.},
        title = "{Infrared spectroscopy of low-mass X-ray binaries - II}",
      journal = {\mnras},
         year = 1999,
        month = jun,
       volume = {306},
       number = {2},
        pages = {417-426},
          doi = {10.1046/j.1365-8711.1999.02547.x},
archivePrefix = {arXiv},
       eprint = {astro-ph/9901327},
 primaryClass = {astro-ph},
       adsurl = {https://ui.adsabs.harvard.edu/abs/1999MNRAS.306..417B}
}

@ARTICLE{2017GeCoA.219...44W,
       author = {{Wang}, Wenzhong and {Zhou}, Chen and {Qin}, Tian and {Kang}, Jin-Ting and {Huang}, Shichun and {Wu}, Zhongqing and {Huang}, Fang},
        title = "{Effect of Ca content on equilibrium Ca isotope fractionation between orthopyroxene and clinopyroxene}",
      journal = {\gca},
         year = 2017,
        month = dec,
       volume = {219},
        pages = {44-56},
          doi = {10.1016/j.gca.2017.09.022},
       adsurl = {https://ui.adsabs.harvard.edu/abs/2017GeCoA.219...44W}
}

@article{ WOS:A1981LC75800001,
Author = {Cameron, M and Papike, JJ},
Title = {STRUCTURAL AND CHEMICAL VARIATIONS IN PYROXENES},
Journal = {AMERICAN MINERALOGIST},
Year = {1981},
Volume = {66},
Number = {1-2},
Pages = {1-50},
ISSN = {0003-004X},
Unique-ID = {WOS:A1981LC75800001},
}

@BOOK{2021cosm.book.....M,
       author = {{McSween}, Harry Y. and {Huss}, Gary R.},
        title = "{Cosmochemistry}",
         year = 2021,
       adsurl = {https://ui.adsabs.harvard.edu/abs/2021cosm.book.....M}
}

@ARTICLE{2019A&A...623A.151P,
       author = {{P{\'e}rez-Mesa}, V. and {Zamora}, O. and {Garc{\'\i}a-Hern{\'a}ndez}, D.~A. and {Osorio}, Y. and {Masseron}, T. and {Plez}, B. and {Manchado}, A. and {Karakas}, A.~I. and {Lugaro}, M.},
        title = "{Exploring circumstellar effects on the lithium and calcium abundances in massive Galactic O-rich AGB stars}",
      journal = {\aap},
         year = 2019,
        month = mar,
       volume = {623},
          eid = {A151},
        pages = {A151},
          doi = {10.1051/0004-6361/201834611},
archivePrefix = {arXiv},
       eprint = {1902.04309},
 primaryClass = {astro-ph.SR},
       adsurl = {https://ui.adsabs.harvard.edu/abs/2019A&A...623A.151P}
}

@ARTICLE{Henning2010,
       author = {{Henning}, Thomas},
        title = "{Cosmic Silicates}",
      journal = {\araa},
         year = 2010,
        month = sep,
       volume = {48},
        pages = {21-46},
          doi = {10.1146/annurev-astro-081309-130815},
       adsurl = {https://ui.adsabs.harvard.edu/abs/2010ARA&A..48...21H}
}

@ARTICLE{2007ApJ...656..615P,
       author = {{Posch}, Th. and {Mutschke}, H. and {Trieloff}, M. and {Henning}, Th.},
        title = "{Infrared Spectroscopy of Calcium-Aluminium-rich Inclusions: Analog Material for Protoplanetary Dust?}",
      journal = {\apj},
         year = 2007,
        month = feb,
       volume = {656},
       number = {1},
        pages = {615-620},
          doi = {10.1086/510445},
       adsurl = {https://ui.adsabs.harvard.edu/abs/2007ApJ...656..615P}
}

@INPROCEEDINGS{2005ASPC..341..605T,
       author = {{Tielens}, A.~G.~G.~M. and {Waters}, L.~B.~F.~M. and {Bernatowicz}, T.~J.},
        title = "{Origin and Evolution of Dust in Circumstellar and Interstellar Environments}",
    booktitle = {Chondrites and the Protoplanetary Disk},
         year = 2005,
       editor = {{Krot}, A.~N. and {Scott}, E.~R.~D. and {Reipurth}, B.},
       series = {Astronomical Society of the Pacific Conference Series},
       volume = {341},
        month = dec,
        pages = {605},
       adsurl = {https://ui.adsabs.harvard.edu/abs/2005ASPC..341..605T}
}

@article{Li2025,
  author  = {Li, C. and Li, Y. and Pang, R. and others},
  title   = {Impact-induced ultra-high melting point oldhamite discovered in {Chang’E}-6 lunar soil},
  journal = {Nature Communications},
  year    = {2025},
  volume  = {16},
  number  = {1},
  pages   = {2155},
  doi     = {10.1038/s41467-025-57337-0},
  url     = {https://doi.org/10.1038/s41467-025-57337-0}
}

@incollection{RUDNICK20141,
title = {4.1 - Composition of the Continental Crust},
editor = {Heinrich D. Holland and Karl K. Turekian},
booktitle = {Treatise on Geochemistry (Second Edition)},
publisher = {Elsevier},
edition = {Second Edition},
address = {Oxford},
pages = {1-51},
year = {2014},
isbn = {978-0-08-098300-4},
doi = {https://doi.org/10.1016/B978-0-08-095975-7.00301-6},
url = {https://www.sciencedirect.com/science/article/pii/B9780080959757003016},
author = {R.L. Rudnick and S. Gao}
}

@ARTICLE{2005ApJ...622..970L,
       author = {{Lee}, Julia C. and {Ravel}, B.},
        title = "{Determining the Grain Composition of the Interstellar Medium with High-Resolution X-Ray Spectroscopy}",
      journal = {\apj},
         year = 2005,
        month = apr,
       volume = {622},
       number = {2},
        pages = {970-976},
          doi = {10.1086/428118},
archivePrefix = {arXiv},
       eprint = {astro-ph/0412393},
 primaryClass = {astro-ph},
       adsurl = {https://ui.adsabs.harvard.edu/abs/2005ApJ...622..970L}
}

@ARTICLE{2020A&A...641A.149R,
       author = {{Rogantini}, D. and {Costantini}, E. and {Zeegers}, S.~T. and {Mehdipour}, M. and {Psaradaki}, I. and {Raassen}, A.~J.~J. and {de Vries}, C.~P. and {Waters}, L.~B.~F.~M.},
        title = "{Magnesium and silicon in interstellar dust: X-ray overview}",
      journal = {\aap},
         year = 2020,
        month = sep,
       volume = {641},
          eid = {A149},
        pages = {A149},
          doi = {10.1051/0004-6361/201936805},
archivePrefix = {arXiv},
       eprint = {2007.03329},
 primaryClass = {astro-ph.HE},
       adsurl = {https://ui.adsabs.harvard.edu/abs/2020A&A...641A.149R}
}

@ARTICLE{2003ExA....16....1W,
       author = {{Weisskopf}, M.~C. and {Aldcroft}, T.~L. and {Bautz}, M. and {Cameron}, R.~A. and {Dewey}, D. and {Drake}, J.~J. and {Grant}, C.~E. and {Marshall}, H.~L. and {Murray}, S.~S.},
        title = "{An Overview of the Performance of the Chandra X-ray Observatory}",
      journal = {Experimental Astronomy},
         year = 2003,
        month = aug,
       volume = {16},
       number = {1},
        pages = {1-68},
          doi = {10.1023/B:EXPA.0000038953.49421.54},
archivePrefix = {arXiv},
       eprint = {astro-ph/0503319},
 primaryClass = {astro-ph},
       adsurl = {https://ui.adsabs.harvard.edu/abs/2003ExA....16....1W}
}

@ARTICLE{2019A&A...630A.143R,
       author = {{Rogantini}, D. and {Costantini}, E. and {Zeegers}, S.~T. and {de Vries}, C.~P. and {Mehdipour}, M. and {de Groot}, F. and {Mutschke}, H. and {Psaradaki}, I. and {Waters}, L.~B.~F.~M.},
        title = "{Interstellar dust along the line of sight of GX 3+1}",
      journal = {\aap},
         year = 2019,
        month = oct,
       volume = {630},
          eid = {A143},
        pages = {A143},
          doi = {10.1051/0004-6361/201935883},
archivePrefix = {arXiv},
       eprint = {1909.00652},
 primaryClass = {astro-ph.GA},
       adsurl = {https://ui.adsabs.harvard.edu/abs/2019A&A...630A.143R}
}

@ARTICLE{1974RvGSP..12...71G,
       author = {{Grossman}, L. and {Larimer}, J.~W.},
        title = "{Early chemical history of the solar system.}",
      journal = {Reviews of Geophysics and Space Physics},
         year = 1974,
        month = jan,
       volume = {12},
        pages = {71-101},
          doi = {10.1029/RG012i001p00071},
       adsurl = {https://ui.adsabs.harvard.edu/abs/1974RvGSP..12...71G}
}

@ARTICLE{1998M&PS...33.1123P,
       author = {{Petaev}, Michail I. and {Wood}, John A.},
        title = "{The condensation with partial isolation model of condensation in the solar nebula}",
      journal = {\maps},
         year = 1998,
        month = sep,
       volume = {33},
       number = {5},
        pages = {1123-1137},
          doi = {10.1111/j.1945-5100.1998.tb01717.x},
       adsurl = {https://ui.adsabs.harvard.edu/abs/1998M&PS...33.1123P}
}

@ARTICLE{2000GeCoA..64..339E,
       author = {{Ebel}, Denton S. and {Grossman}, Lawrence},
        title = "{Condensation in dust-enriched systems}",
      journal = {\gca},
         year = 2000,
        month = jan,
       volume = {64},
       number = {2},
        pages = {339-366},
          doi = {10.1016/S0016-7037(99)00284-7},
       adsurl = {https://ui.adsabs.harvard.edu/abs/2000GeCoA..64..339E}
}

@ARTICLE{1995GeCoA..59.3413Y,
       author = {{Yoneda}, Shigekazu and {Grossman}, Lawrence},
        title = "{Condensation of CaO sbnd MgO sbnd Al $_{2}$O $_{3}$sbnd SiO $_{2}$ liquids from cosmic gases}",
      journal = {\gca},
         year = 1995,
        month = aug,
       volume = {59},
       number = {16},
        pages = {3413-3444},
          doi = {10.1016/0016-7037(95)00214-K},
       adsurl = {https://ui.adsabs.harvard.edu/abs/1995GeCoA..59.3413Y}
}

@ARTICLE{1993GeCoA..57.2377W,
       author = {{Wood}, J.~A. and {Hashimoto}, A.},
        title = "{Mineral equilibrium in fractionated nebular systems}",
      journal = {\gca},
         year = 1993,
        month = may,
       volume = {57},
       number = {10},
        pages = {2377-2388},
          doi = {10.1016/0016-7037(93)90575-H},
       adsurl = {https://ui.adsabs.harvard.edu/abs/1993GeCoA..57.2377W}
}

@ARTICLE{2025ApJ...986...41R,
       author = {{Rogantini}, Daniele and {Homan}, Jeroen and {Plotkin}, Richard M. and {van den Berg}, Maureen and {Miller-Jones}, James and {Neilsen}, Joey and {Chakrabarty}, Deepto and {Fender}, Rob P. and {Schulz}, Norbert},
        title = "{A Persistent Disk Wind and Variable Jet Outflow in the Neutron-star Low-mass X-Ray Binary GX 13+1}",
      journal = {\apj},
         year = 2025,
        month = jun,
       volume = {986},
       number = {1},
          eid = {41},
        pages = {41},
          doi = {10.3847/1538-4357/adcabe},
archivePrefix = {arXiv},
       eprint = {2504.05452},
 primaryClass = {astro-ph.HE},
       adsurl = {https://ui.adsabs.harvard.edu/abs/2025ApJ...986...41R}
}

@ARTICLE{2011AJ....141..129H,
       author = {{Huenemoerder}, David P. and {Mitschang}, Arik and {Dewey}, Daniel and {Nowak}, Michael A. and {Schulz}, Norbert S. and {Nichols}, Joy S. and {Davis}, John E. and {Houck}, John C. and {Marshall}, Herman L. and {Noble}, Michael S. and {Morgan}, Doug and {Canizares}, Claude R.},
        title = "{TGCat: The Chandra Transmission Grating Data Catalog and Archive}",
      journal = {\aj},
         year = 2011,
        month = apr,
       volume = {141},
       number = {4},
          eid = {129},
        pages = {129},
          doi = {10.1088/0004-6256/141/4/129},
       adsurl = {https://ui.adsabs.harvard.edu/abs/2011AJ....141..129H}
}

@INPROCEEDINGS{2006SPIE.6270E..1VF,
       author = {{Fruscione}, Antonella and {McDowell}, Jonathan C. and {Allen}, Glenn E. and {Brickhouse}, Nancy S. and {Burke}, Douglas J. and {Davis}, John E. and {Durham}, Nick and {Elvis}, Martin and {Galle}, Elizabeth C. and {Harris}, Daniel E. and {Huenemoerder}, David P. and {Houck}, John C. and {Ishibashi}, Bish and {Karovska}, Margarita and {Nicastro}, Fabrizio and {Noble}, Michael S. and {Nowak}, Michael A. and {Primini}, Frank A. and {Siemiginowska}, Aneta and {Smith}, Randall K. and {Wise}, Michael},
        title = "{CIAO: Chandra's data analysis system}",
    booktitle = {Observatory Operations: Strategies, Processes, and Systems},
         year = 2006,
       editor = {{Silva}, David R. and {Doxsey}, Rodger E.},
       series = {Society of Photo-Optical Instrumentation Engineers (SPIE) Conference Series},
       volume = {6270},
        month = jun,
          eid = {62701V},
        pages = {62701V},
          doi = {10.1117/12.671760},
       adsurl = {https://ui.adsabs.harvard.edu/abs/2006SPIE.6270E..1VF}
}

@ARTICLE{2025PASJ...77S...1T,
       author = {{Tashiro}, Makoto and {Kelley}, Richard and {Watanabe}, Shin and {Maejima}, Hironori and {Reichenthal}, Lillian and {Toda}, Kenichi and {Hartz}, Leslie and {Santovincenzo}, Andrea and {Matsushita}, Kyoko and {Yamaguchi}, Hiroya and {Petre}, Robert and {Williams}, Brian and {Guainazzi}, Matteo and {Costantini}, Elisa and {Takei}, Yoh and {Ishisaki}, Yoshitaka and {Fujimoto}, Ryuichi and {Henegar-Leon}, Joy and {Sneiderman}, Gary and {Tomida}, Hiroshi and {Mori}, Koji and {Nakajima}, Hiroshi and {Terada}, Yukikatsu and {Holland}, Matthew and {Loewenstein}, Michael and {Miller}, Eric and {Sawada}, Makoto and {Kallman}, Timothy and {Kaastra}, Jelle and {Done}, Chris and {Enoto}, Teruaki and {Bamba}, Aya and {Corrales}, Lia and {Ueda}, Yoshihiro and {Kara}, Erin and {Zhuravleva}, Irina and {Fujita}, Yutaka and {Arai}, Yoshitaka and {Audard}, Marc and {Awaki}, Hisamitsu and {Ballhausen}, Ralf and {Baluta}, Chris and {Bando}, Nobutaka and {Behar}, Ehud and {Bialas}, Thomas and {Boissay-Malaquin}, Rozenn and {Brenneman}, Laura and {Brown}, Gregory V. and {Chiao}, Meng and {Cumbee}, Renata and {de Vries}, Cor and {den Herder}, Jan-Willem and {D{\'\i}az Trigo}, Mar{\'\i}a and {DiPirro}, Michael and {Dotani}, Tadayasu and {Carrero}, Jacobo Ebrero and {Ebisawa}, Ken and {Eckart}, Megan and {Eckert}, Dominique and {Eguchi}, Satoshi and {Ezoe}, Yuichiro and {Ferrigno}, Carlo and {Foster}, Adam and {Fukazawa}, Yasushi and {Fukushima}, Kotaro and {Furuzawa}, Akihiro and {Gallo}, Luigi C. and {Garcia Martinez}, Javier and {Gorter}, Nathalie and {Grim}, Martin and {Gu}, Liyi and {Hagino}, Kouichi and {Hamaguchi}, Kenji and {Hatsukade}, Isamu and {Hayashi}, Katsuhiro and {Hayashi}, Takayuki and {Hell}, Natalie and {Hodges-Kluck}, Edmund and {Horiuchi}, Takafumi and {Hornschemeier}, Ann and {Hoshino}, Akio and {Ichinohe}, Yuto and {Ikuta}, Chisato and {Iizuka}, Ryo and {Ishi}, Daiki and {Ishida}, Manabu and {Ishihama}, Naoki and {Ishikawa}, Kumi and {Ishimura}, Kosei and {Jaffe}, Tess and {Katsuda}, Satoru and {Kanemaru}, Yoshiaki and {Kenyon}, Steven and {Kilbourne}, Caroline and {Kimball}, Mark and {Kitamoto}, Shunji and {Kobayashi}, Shogo and {Kohmura}, Takayoshi and {Kubota}, Aya and {Leutenegger}, Maurice A. and {Maeda}, Yoshitomo and {Markevitch}, Maxim and {Matsumoto}, Hironori and {Matsuzaki}, Keiichi and {McCammon}, Dan and {McLaughlin}, Brian and {McNamara}, Brian and {Mernier}, Fran{\c{c}}ois and {Miko}, Joseph and {Miller}, Jon M. and {Minesugi}, Kenji and {Mitani}, Shinji and {Mitsuishi}, Ikuyuki and {Mizumoto}, Misaki and {Mizuno}, Tsunefumi and {Mukai}, Koji and {Murakami}, Hiroshi and {Mushotzky}, Richard and {Nakazawa}, Kazuhiro and {Natsukari}, Chikara and {Ness}, Jan-Uwe and {Nigo}, Kenichiro and {Nishiyama}, Mari and {Nobukawa}, Kumiko and {Nobukawa}, Masayoshi and {Noda}, Hirofumi and {Odaka}, Hirokazu and {Ogawa}, Mina and {Ogawa}, Shoji and {Ogorzalek}, Anna and {Okajima}, Takashi and {Okamoto}, Atsushi and {Ota}, Naomi and {Ozaki}, Masanobu and {Paltani}, Stephane and {Plucinsky}, Paul and {Porter}, F. Scott and {Pottschmidt}, Katja and {Quero}, Jose Antonio and {Sasaki}, Takahiro and {Sato}, Kosuke and {Sato}, Rie and {Sato}, Toshiki and {Sato}, Yoichi and {Seta}, Hiromi and {Shida}, Maki and {Shidatsu}, Megumi and {Shigeto}, Shuhei and {Shipman}, Russel and {Shinozaki}, Keisuke and {Shirron}, Peter and {Simionescu}, Aurora and {Smith}, Randall K. and {Soong}, Yang and {Suzuki}, Hiromasa and {Szymkowiak}, Andrew and {Takahashi}, Hiromitsu and {Takeo}, Mai and {Tamagawa}, Toru and {Tamura}, Keisuke and {Tanaka}, Takaaki and {Tanimoto}, Atsushi and {Terashima}, Yuichi and {Tsuboi}, Yohko and {Tsujimoto}, Masahiro and {Tsunemi}, Hiroshi and {Tsuru}, Takeshi Go and {Uchida}, Hiroyuki and {Uchida}, Nagomi and {Uchida}, Yuusuke and {Uchiyama}, Hideki and {Uno}, Shinichiro and {Vink}, Jacco and {Witthoeft}, Michael and {Wolfs}, Rob and {Yamada}, Satoshi and {Yamada}, Shinya and {Yamaoka}, Kazutaka and {Yamasaki}, Noriko and {Yamauchi}, Makoto and {Yamauchi}, Shigeo and {Yanagase}, Keiichi and {Yaqoob}, Tahir and {Yasuda}, Susumu and {Yoneyama}, Tomokage and {Yoshida}, Tessei and {Yukita}, Mihoko},
        title = "{X-Ray Imaging and Spectroscopy Mission}",
      journal = {\pasj},
         year = 2025,
        month = sep,
       volume = {77},
        pages = {S1-S9},
          doi = {10.1093/pasj/psaf023},
       adsurl = {https://ui.adsabs.harvard.edu/abs/2025PASJ...77S...1T}
}

@ARTICLE{Psaradaki2020,
       author = {{Psaradaki}, I. and {Costantini}, E. and {Mehdipour}, M. and {Rogantini}, D. and {de Vries}, C.~P. and {de Groot}, F. and {Mutschke}, H. and {Trasobares}, S. and {Waters}, L.~B.~F.~M. and {Zeegers}, S.~T.},
        title = "{Interstellar oxygen along the line of sight of Cygnus X-2}",
      journal = {\aap},
         year = 2020,
        month = oct,
       volume = {642},
          eid = {A208},
        pages = {A208},
          doi = {10.1051/0004-6361/202038749},
archivePrefix = {arXiv},
       eprint = {2009.06244},
 primaryClass = {astro-ph.HE},
       adsurl = {https://ui.adsabs.harvard.edu/abs/2020A&A...642A.208P}
}

@ARTICLE{Tashiro2025,
       author = {{Tashiro}, Makoto and {Kelley}, Richard and {Watanabe}, Shin and {Maejima}, Hironori and {Reichenthal}, Lillian and {Toda}, Kenichi and {Hartz}, Leslie and {Santovincenzo}, Andrea and {Matsushita}, Kyoko and {Yamaguchi}, Hiroya and {Petre}, Robert and {Williams}, Brian and {Guainazzi}, Matteo and {Costantini}, Elisa and {Takei}, Yoh and {Ishisaki}, Yoshitaka and {Fujimoto}, Ryuichi and {Henegar-Leon}, Joy and {Sneiderman}, Gary and {Tomida}, Hiroshi and {Mori}, Koji and {Nakajima}, Hiroshi and {Terada}, Yukikatsu and {Holland}, Matthew and {Loewenstein}, Michael and {Miller}, Eric and {Sawada}, Makoto and {Kallman}, Timothy and {Kaastra}, Jelle and {Done}, Chris and {Enoto}, Teruaki and {Bamba}, Aya and {Corrales}, Lia and {Ueda}, Yoshihiro and {Kara}, Erin and {Zhuravleva}, Irina and {Fujita}, Yutaka and {Arai}, Yoshitaka and {Audard}, Marc and {Awaki}, Hisamitsu and {Ballhausen}, Ralf and {Baluta}, Chris and {Bando}, Nobutaka and {Behar}, Ehud and {Bialas}, Thomas and {Boissay-Malaquin}, Rozenn and {Brenneman}, Laura and {Brown}, Gregory V. and {Chiao}, Meng and {Cumbee}, Renata and {de Vries}, Cor and {den Herder}, Jan-Willem and {D{\'\i}az Trigo}, Mar{\'\i}a and {DiPirro}, Michael and {Dotani}, Tadayasu and {Carrero}, Jacobo Ebrero and {Ebisawa}, Ken and {Eckart}, Megan and {Eckert}, Dominique and {Eguchi}, Satoshi and {Ezoe}, Yuichiro and {Ferrigno}, Carlo and {Foster}, Adam and {Fukazawa}, Yasushi and {Fukushima}, Kotaro and {Furuzawa}, Akihiro and {Gallo}, Luigi C. and {Garcia Martinez}, Javier and {Gorter}, Nathalie and {Grim}, Martin and {Gu}, Liyi and {Hagino}, Kouichi and {Hamaguchi}, Kenji and {Hatsukade}, Isamu and {Hayashi}, Katsuhiro and {Hayashi}, Takayuki and {Hell}, Natalie and {Hodges-Kluck}, Edmund and {Horiuchi}, Takafumi and {Hornschemeier}, Ann and {Hoshino}, Akio and {Ichinohe}, Yuto and {Ikuta}, Chisato and {Iizuka}, Ryo and {Ishi}, Daiki and {Ishida}, Manabu and {Ishihama}, Naoki and {Ishikawa}, Kumi and {Ishimura}, Kosei and {Jaffe}, Tess and {Katsuda}, Satoru and {Kanemaru}, Yoshiaki and {Kenyon}, Steven and {Kilbourne}, Caroline and {Kimball}, Mark and {Kitamoto}, Shunji and {Kobayashi}, Shogo and {Kohmura}, Takayoshi and {Kubota}, Aya and {Leutenegger}, Maurice A. and {Maeda}, Yoshitomo and {Markevitch}, Maxim and {Matsumoto}, Hironori and {Matsuzaki}, Keiichi and {McCammon}, Dan and {McLaughlin}, Brian and {McNamara}, Brian and {Mernier}, Fran{\c{c}}ois and {Miko}, Joseph and {Miller}, Jon M. and {Minesugi}, Kenji and {Mitani}, Shinji and {Mitsuishi}, Ikuyuki and {Mizumoto}, Misaki and {Mizuno}, Tsunefumi and {Mukai}, Koji and {Murakami}, Hiroshi and {Mushotzky}, Richard and {Nakazawa}, Kazuhiro and {Natsukari}, Chikara and {Ness}, Jan-Uwe and {Nigo}, Kenichiro and {Nishiyama}, Mari and {Nobukawa}, Kumiko and {Nobukawa}, Masayoshi and {Noda}, Hirofumi and {Odaka}, Hirokazu and {Ogawa}, Mina and {Ogawa}, Shoji and {Ogorzalek}, Anna and {Okajima}, Takashi and {Okamoto}, Atsushi and {Ota}, Naomi and {Ozaki}, Masanobu and {Paltani}, Stephane and {Plucinsky}, Paul and {Porter}, F. Scott and {Pottschmidt}, Katja and {Quero}, Jose Antonio and {Sasaki}, Takahiro and {Sato}, Kosuke and {Sato}, Rie and {Sato}, Toshiki and {Sato}, Yoichi and {Seta}, Hiromi and {Shida}, Maki and {Shidatsu}, Megumi and {Shigeto}, Shuhei and {Shipman}, Russel and {Shinozaki}, Keisuke and {Shirron}, Peter and {Simionescu}, Aurora and {Smith}, Randall K. and {Soong}, Yang and {Suzuki}, Hiromasa and {Szymkowiak}, Andrew and {Takahashi}, Hiromitsu and {Takeo}, Mai and {Tamagawa}, Toru and {Tamura}, Keisuke and {Tanaka}, Takaaki and {Tanimoto}, Atsushi and {Terashima}, Yuichi and {Tsuboi}, Yohko and {Tsujimoto}, Masahiro and {Tsunemi}, Hiroshi and {Tsuru}, Takeshi Go and {Uchida}, Hiroyuki and {Uchida}, Nagomi and {Uchida}, Yuusuke and {Uchiyama}, Hideki and {Uno}, Shinichiro and {Vink}, Jacco and {Witthoeft}, Michael and {Wolfs}, Rob and {Yamada}, Satoshi and {Yamada}, Shinya and {Yamaoka}, Kazutaka and {Yamasaki}, Noriko and {Yamauchi}, Makoto and {Yamauchi}, Shigeo and {Yanagase}, Keiichi and {Yaqoob}, Tahir and {Yasuda}, Susumu and {Yoneyama}, Tomokage and {Yoshida}, Tessei and {Yukita}, Mihoko},
        title = "{X-Ray Imaging and Spectroscopy Mission}",
      journal = {\pasj},
         year = 2025,
        month = sep,
       volume = {77},
        pages = {S1-S9},
          doi = {10.1093/pasj/psaf023},
       adsurl = {https://ui.adsabs.harvard.edu/abs/2025PASJ...77S...1T}
}

@INCOLLECTION{Costantini2022,
       author = {{Costantini}, E. and {Corrales}, L.},
        title = "{Interstellar Absorption and Dust Scattering}",
    booktitle = {Handbook of X-ray and Gamma-ray Astrophysics},
         year = 2022,
       editor = {{Bambi}, Cosimo and {Sangangelo}, Andrea},
          eid = {40},
        pages = {40},
          doi = {10.1007/978-981-16-4544-0_93-1},
       adsurl = {https://ui.adsabs.harvard.edu/abs/2022hxga.book...40C}
}

@Article{Hunter:2007,
  Author    = {Hunter, J. D.},
  Title     = {Matplotlib: A 2D graphics environment},
  Journal   = {Computing in Science \& Engineering},
  Volume    = {9},
  Number    = {3},
  Pages     = {90--95},
  publisher = {IEEE COMPUTER SOC},
  doi       = {10.1109/MCSE.2007.55},
  year      = 2007
}

@Article{harris2020array,
 title         = {Array programming with {NumPy}},
 author        = {Charles R. Harris and K. Jarrod Millman and St{\'{e}}fan J.
                 van der Walt and Ralf Gommers and Pauli Virtanen and David
                 Cournapeau and Eric Wieser and Julian Taylor and Sebastian
                 Berg and Nathaniel J. Smith and Robert Kern and Matti Picus
                 and Stephan Hoyer and Marten H. van Kerkwijk and Matthew
                 Brett and Allan Haldane and Jaime Fern{\'{a}}ndez del
                 R{\'{i}}o and Mark Wiebe and Pearu Peterson and Pierre
                 G{\'{e}}rard-Marchant and Kevin Sheppard and Tyler Reddy and
                 Warren Weckesser and Hameer Abbasi and Christoph Gohlke and
                 Travis E. Oliphant},
 year          = {2020},
 month         = sep,
 journal       = {Nature},
 volume        = {585},
 number        = {7825},
 pages         = {357--362},
 doi           = {10.1038/s41586-020-2649-2},
 publisher     = {Springer Science and Business Media {LLC}},
 url           = {https://doi.org/10.1038/s41586-020-2649-2}
}

@ARTICLE{2014E&PSL.394..135V,
       author = {{Valdes}, Maria C. and {Moreira}, Manuel and {Foriel}, Julien and {Moynier}, Fr{\'e}d{\'e}ric},
        title = "{The nature of Earth's building blocks as revealed by calcium isotopes}",
      journal = {Earth and Planetary Science Letters},
         year = 2014,
        month = may,
       volume = {394},
        pages = {135-145},
          doi = {10.1016/j.epsl.2014.02.052},
       adsurl = {https://ui.adsabs.harvard.edu/abs/2014E&PSL.394..135V}
}

@ARTICLE{2016ApJ...827...49S,
       author = {{Schulz}, Norbert S. and {Corrales}, Lia and {Canizares}, Claude R.},
        title = "{Si K Edge Structure and Variability in Galactic X-Ray Binaries}",
      journal = {\apj},
         year = 2016,
        month = aug,
       volume = {827},
       number = {1},
          eid = {49},
        pages = {49},
          doi = {10.3847/0004-637X/827/1/49},
archivePrefix = {arXiv},
       eprint = {1605.04837},
 primaryClass = {astro-ph.GA},
       adsurl = {https://ui.adsabs.harvard.edu/abs/2016ApJ...827...49S}
}

@incollection{BREARLEY2003247,
title = {1.09 - Nebular versus Parent-body Processing},
editor = {Heinrich D. Holland and Karl K. Turekian},
booktitle = {Treatise on Geochemistry},
publisher = {Pergamon},
address = {Oxford},
pages = {247-268},
year = {2003},
isbn = {978-0-08-043751-4},
doi = {https://doi.org/10.1016/B0-08-043751-6/01068-9},
url = {https://www.sciencedirect.com/science/article/pii/B0080437516010689},
author = {A.J. Brearley}
}

@ARTICLE{1992GeCoA..56.2873M,
       author = {{Metzler}, K. and {Bischoff}, A. and {Stoeffler}, D.},
        title = "{Accretionary dust mantles in CM chondrites: Evidence for solar nebula processes}",
      journal = {\gca},
         year = 1992,
        month = jul,
       volume = {56},
       number = {7},
        pages = {2873-2897},
          doi = {10.1016/0016-7037(92)90365-P},
       adsurl = {https://ui.adsabs.harvard.edu/abs/1992GeCoA..56.2873M}
}

@ARTICLE{2025SSRv..221...11L,
       author = {{Lee}, Martin R. and {Alexander}, Conel M. O'D. and {Bischoff}, Addi and {Brearley}, Adrian J. and {Dobric{\u{a}}}, Elena and {Fujiya}, Wataru and {Le Guillou}, Corentin and {King}, Ashley J. and {van Kooten}, Elishevah and {Krot}, Alexander N. and {Leitner}, Jan and {Marrocchi}, Yves and {Patzek}, Markus and {Petaev}, Michail I. and {Piani}, Laurette and {Pravdivtseva}, Olga and {Remusat}, Laurent and {Telus}, Myriam and {Tsuchiyama}, Akira and {Vacher}, Lionel G.},
        title = "{Low-Temperature Aqueous Alteration of Chondrites}",
      journal = {\ssr},
         year = 2025,
        month = feb,
       volume = {221},
       number = {1},
          eid = {11},
        pages = {11},
          doi = {10.1007/s11214-024-01132-8},
       adsurl = {https://ui.adsabs.harvard.edu/abs/2025SSRv..221...11L}
}

@ARTICLE{2017A&A...599A.117Z,
       author = {{Zeegers}, S.~T. and {Costantini}, E. and {de Vries}, C.~P. and {Tielens}, A.~G.~G.~M. and {Chihara}, H. and {de Groot}, F. and {Mutschke}, H. and {Waters}, L.~B.~F.~M. and {Zeidler}, S.},
        title = "{Absorption and scattering by interstellar dust in the silicon K-edge of GX 5-1}",
      journal = {\aap},
         year = 2017,
        month = mar,
       volume = {599},
          eid = {A117},
        pages = {A117},
          doi = {10.1051/0004-6361/201628507},
archivePrefix = {arXiv},
       eprint = {1612.07988},
 primaryClass = {astro-ph.GA},
       adsurl = {https://ui.adsabs.harvard.edu/abs/2017A&A...599A.117Z}
}

@ARTICLE{Cardelli1991,
       author = {{Cardelli}, Jason A. and {Federman}, S.~R. and {Smith}, V.~V.},
        title = "{Interstellar Environments Probed by CA i Absorption and the Effects of Density-dependent Depletions}",
      journal = {\apjl},
         year = 1991,
        month = nov,
       volume = {381},
        pages = {L17},
          doi = {10.1086/186186},
       adsurl = {https://ui.adsabs.harvard.edu/abs/1991ApJ...381L..17C}
}

@ARTICLE{Edgar1989,
       author = {{Edgar}, Richard J. and {Savage}, Blair D.},
        title = "{The Density Distribution of Refractory Elements Away from the Galactic Plane}",
      journal = {\apj},
         year = 1989,
        month = may,
       volume = {340},
        pages = {762},
          doi = {10.1086/167435},
       adsurl = {https://ui.adsabs.harvard.edu/abs/1989ApJ...340..762E}
}

@ARTICLE{Phillips1984,
       author = {{Phillips}, A.~P. and {Pettini}, M. and {Gondhalekar}, P.~M.},
        title = "{Element depletions in interstellar gas - II. The density-dependence of calcium and sodium depletions.}",
      journal = {\mnras},
         year = 1984,
        month = jan,
       volume = {206},
        pages = {337-350},
          doi = {10.1093/mnras/206.2.337},
       adsurl = {https://ui.adsabs.harvard.edu/abs/1984MNRAS.206..337P}
}

@ARTICLE{Welty1996,
       author = {{Welty}, Daniel E. and {Morton}, Donald C. and {Hobbs}, L.~M.},
        title = "{A High-Resolution Survey of Interstellar Ca II Absorption}",
      journal = {\apjs},
         year = 1996,
        month = oct,
       volume = {106},
        pages = {533},
          doi = {10.1086/192347},
       adsurl = {https://ui.adsabs.harvard.edu/abs/1996ApJS..106..533W}
}

@ARTICLE{1995A&AS..109..125V,
       author = {{Verner}, D.~A. and {Yakovlev}, D.~G.},
        title = "{Analytic FITS for partial photoionization cross sections.}",
      journal = {\aaps},
         year = 1995,
        month = jan,
       volume = {109},
        pages = {125-133},
       adsurl = {https://ui.adsabs.harvard.edu/abs/1995A&AS..109..125V}
}

@ARTICLE{Kallman2019,
       author = {{Kallman}, T. and {McCollough}, M. and {Koljonen}, K. and {Liedahl}, D. and {Miller}, J. and {Paerels}, F. and {Pooley}, G. and {Sako}, M. and {Schulz}, N. and {Trushkin}, S. and {Corrales}, L.},
        title = "{Photoionization Emission Models for the Cyg X-3 X-Ray Spectrum}",
      journal = {\apj},
         year = 2019,
        month = mar,
       volume = {874},
       number = {1},
          eid = {51},
        pages = {51},
          doi = {10.3847/1538-4357/ab09f8},
archivePrefix = {arXiv},
       eprint = {1902.05589},
 primaryClass = {astro-ph.HE},
       adsurl = {https://ui.adsabs.harvard.edu/abs/2019ApJ...874...51K}
}
\bibliographystyle{aasjournal}
\newpage

\end{document}